\documentclass[10pt,twoside,leqno]{article}
\usepackage{amsfonts}
\usepackage{epsfig}
\usepackage{subfig}
\usepackage{amssymb,amsmath,bm}
\usepackage{amsthm}
\usepackage{mathrsfs}
\usepackage[section]{algorithm}
\usepackage{fancyhdr}
\theoremstyle{plain}

\newcounter{hypA}

\begin{document}

\bigskip

\begin{center}

{\Large \textbf{A Bayesian Mixture of Lasso Regressions with $t-$Errors}}

\bigskip

BY ALBERTO COZZINI$^{1}$, AJAY JASRA$^{2}$ \& GIOVANNI MONTANA$^{1}$

{\footnotesize $^{1}$Department of Mathematics,
Imperial College London, London, SW7 2AZ, UK.}\\
{\footnotesize E-Mail:\,}\texttt{\emph{\footnotesize a.m.cozzini@ic.ac.uk, g.montana@ic.ac.uk}}\\
{\footnotesize $^{2}$Department of Statistics \& Applied Probability,
National University of Singapore, Singapore, 117546, Sg.}\\
{\footnotesize E-Mail:\,}\texttt{\emph{\footnotesize staja@nus.edu.sg}}
\end{center}

\begin{abstract}
%We are presented with historical financial price data from a variety of financial markets. From the perspective of a systematic (or automated) trader the macro sector to which a product belongs to, is %less useful as one would like to cluster financial instruments with similar price dynamics. Using the profitability of a simple trading strategy as a response and a variety of summary statistics from %the prices dynamics as predictors we have heterogeneous populations of regression data. 
Motivated by a challenging problem in financial trading 
we are presented with a mixture of regressions with variable selection problem. In this regard, one is faced with data which possess outliers, skewness and, simultaneously, due to the nature of financial trading, one would like to be able to construct clusters with specific predictors that are fairly sparse. We develop a Bayesian mixture of lasso regressions with $t-$errors to reflect these specific demands. The resulting model is necessarily complex and to fit the model to real data, we develop a state-of-the-art Particle Markov chain Monte Carlo (PMCMC) algorithm based upon sequential Monte Carlo (SMC) methods. The model and algorithm are investigated on both simulated and real data.
\\
\begin{small}
\emph{Some Key Words}:  
Mixture of Regressions, Variable Selection, Particle Markov chain Monte Carlo
\end{small}
\end{abstract}

\section{Introduction}

In the following article, we will consider a Bayesian mixture of lasso regressions with $t-$errors that is motivated by a particular problem in finance. The specifics of the
data are explained in Section \ref{sec:data}, but the model and resulting MCMC algorithm are generic and hence we consider a general presentation during the article.
The data we are presented with is a collection of $n \in \mathbb{N}^{+}$ paired observations $\mathcal{D}_n=(\bm{x}_i,y_i)_{i=1}^{n}$ where $y_i \in \mathbb {R}$ is the response variable and $\bm{x}_i \in \mathbb {R}^{p}$ is the corresponding vector of explanatory variables.
The specific objective is to cluster linear regression curves which satisfy the following constraints:
\begin{itemize}
\item{The regression curves are resistant to outliers}
\item{Each regression curve is specific to each cluster, in that the predictors for one curve may not be present in another}
\item{One would like to have relatively few predictors in each curve}
\end{itemize}
It is remarked that whilst we have been motivated by a problem in finance, this particular scenario is present in other real problems, such as gene expression data; see for example Cozzini et al.~(2011).

In the context of the problem, we will then consider a mixture of regressions, from the Bayesian perspective. Mixture of regressions has been well studied; see for example Goldfield \& Quandt (1973) and Hurn et al.~(2003). 
In addition, the issue of variable selection has also been substantially investigated, both in the supervised and unsupervised mixture modelling setting, by Raftery \& Dean (2006) and Yau \& Holmes (2011), for example.
However,
to our knowledge, there are very few articles which develop a (Bayesian)  model with component specific variable selection, which we incorporate into our model.
An exception in the case of frequentist statistics is Khalili \& Chen (2007). 
We remark that we do not address the issue of selecting the number of components in the mixture, but this is discussed in Section \ref{sec:summary}.
To deal with the issues of robustness to outliers and sparseness of solutions, we consider well-known procedures, by incorporating heavy-tailed regression error as well as a Bayesian lasso type structure (e.g.~Park \& Casella (2008)).
The latter idea has also been followed by Yau \& Holmes (2011). 

These model components lead to a Bayesian statistical model which is very high-dimensional. In order to draw statistical inference, after marginalization, we are left with a posterior distribution on the class labels of the mixture, component specific variable selection indicators and some
additional parameters. Due to the complexity of the resulting posterior, very sophisticated computational tools are required. We focus on using PMCMC (Andrieu et al.~2010), which is particularly useful for statistical models with latent variables.
The PMCMC algorithm uses an SMC algorithm (e.g.~Doucet et al.~(2001)) to update latent variables: we focus on the class labels which have a larger state-space than the variable selection in the examples considered. We develop an SMC algorithm
and subsequently a conditional SMC algorithm for our particle Gibbs algorithm (a special case of PMCMC). The PMCMC algorithm reflects the current state-of-the-art in Bayesian computation and gives us the best chance of reliable inference from the posterior;
although we remark that it is far from infallible and can break down for sufficiently complex problems.

The outline of the paper is as follows.
 In Section \ref{sec:regmdl} we describe the hierarchical representation of the model and justify the choice of priors that lead to the posteriors of interest. In Section \ref{sec:regmeth} we present our PMCMC algorithm. 
In Section \ref{sec:exres} we investigate the model and algorithm on simulated data. 
In Section \ref{sec:data} we describe the applied problem and the data which we analyze.
In Section \ref{sec:summary} the article is concluded and some
avenues of future work are discussed.

%\section{The Data}
\section{The Model}\label{sec:regmdl}

\subsection{Set-Up}

Generalising the peculiarities of the financial data we want to investigate, let us first highlight the relevant aspects of the problem that motivate the mixture of regression model we propose.

Recall we have a collection of $n \in \mathbb{N}^{+}$ paired observations $\mathcal{D}_n=(\bm{x}_i,y_i)_{i=1}^{n}$ where $y_i \in \mathbb {R}$ is the response variable and $\bm{x}_i \in \mathbb {R}^{p}$ is the corresponding vector of explanatory variables. To simplify notation we use $x_{1:p,i}$ to indicate the collection of covariates at the $i^{th}$ sample and set the first element ${x}_{1,i}$ to be $1$ to allow a more convenient formulation of the model. The defining characteristic of the data is that the $n$ samples are generated from a heterogeneous population and only few of the $p$ covariates convey any useful information to explain the variability of the $y_i$. 

To answer these demanding conditions, we propose a Bayesian mixture model which postulates that there are $K \leq n$ possible linear regression curves (one can consider more general basis function, but this is not done here) to describe the data and that each curve potentially depends upon a different collection of the variables $1,\dots,p$. To facilitate the derivation of a sparse solution, we introduce a $p$-dimensional binary vector {$\gamma_{1:p}^k$}, where we use $\gamma^k_{1:p}$ to denote $(\gamma^k_1, \ldots, \gamma^k_p)$, which encodes whether each of $p$ observed covariates should be included or not in the $k^{th}$ regression curve for $k=1, \ldots, K$. Similarly, we use ${\gamma_{1:p}^k}$ as a subscript indicator which deletes the elements corresponding to $\gamma_d^k=0$ for $d \in \{1,\ldots,p\}$ and returns a vector of length $|\gamma_{1:p}^k|_1$ ($\mathbb{L}_1-$norm).

The mixture model is then defined as the conditional distribution of $y_i$ given $\bm{x}_i$
\begin{equation}\label{mixlasso}
y_i | \bm{x}_i, \, \bm{\beta}, \,  \bm{w}, \, \bm{s}_i, \, \bm{\gamma}_{1:p} \, \sim \sum_{k=1}^{K} w_k \, \mathcal{N}( x'_{\gamma_{1:p}^k,i} \, \beta_{\gamma_{1:p}^k}^k , s_i^k)
\end{equation}
where $\mathcal{N}_l(\mu,\Sigma)$ is the $l-$dimensional normal distribution of mean $\mu$ and covariance $\Sigma$. Note that, to simplify notation, when $l=1$ we drop the subscript. 
\eqref{mixlasso} is a mixture of normal distributions with parameters 
\begin{itemize}

\item{$w_k$} with $0 \leq w_k \leq 1 $  for $k=\{1,\ldots,K\}$ such that $\sum_{k=1}^{K}\,w_k=1$, are the mixing proportion of the $K$ components.

\item{$\beta_{1:p}^k$} with $\beta_d^k \in \mathbb{R}$ for $d=\{1,\ldots,p\}$, is the collection of regression coefficients.

\item{${s}^k_{i}$}, with $s_i^k\in \mathbb{R}^+$ for $i=\{1,\ldots,n\}$ is a variable introduced to allow a Student $t-$regression error.

\end{itemize}

Having defined the model, the values of the parameters $\bm{\Psi}=( \bm{w}, \,\bm{\beta}, \, \bm{s}, \, \bm{\gamma})$ are unknown and will have to be inferred from the data $\mathcal{D}_n$ using a Bayesian approach.

Note that throughout our discussion we assume that the number of clusters $K$ is known. In a different situation, we could have included $K$ in the set of unknown parameters and modified the estimation process accordingly. While this would be a standard procedure, it adds another level of complexity to the model that we rather avoid here since it is not the focus of our investigation; see Section \ref{sec:summary} for some discussion.

%------------------------------------------
\subsection {Hierarchical Specification}

Whilst a mixture of Gaussian distributions as described in \eqref{mixlasso} is a fairly general model, it is also flexible enough to allow us to choose convenient priors that achieve the objective of making the model robust to outliers and selecting only the relevant covariates. This task is facilitated by using a hierarchical representation of the mixture model and having different levels of priors and hyperpriors.

Following the standard missing data approach, see Diebolt \& Robert (1994), we introduce, for every $i^{th}-$data point, the latent allocation variable $z_i\in\{1,\dots,K\}$ which indicates the membership of $y_i$ to the $k^{th}-$cluster. Thus, we can simplify the mixture structure and note that the conditional distribution of $y_i$ given $z_i=k$, with probability $p(z_i=k) =  w_k$, is the Gaussian distribution
\begin{equation} \label{condlasso}
y_i  \, | \, \gamma_{1:p}^k, \, x_{\gamma_{1:p}^k}^k, \, \beta_{\gamma_{1:p}^k}^k, \, s^k_i, \, z_i=k \sim  \mathcal{N}( x'_{\gamma_{1:p}^k,i} \, \beta_{\gamma_{1:p}^k}^k , s_i^k).
\end{equation}
Assuming the mixture weights follow a Dirichlet distribution, the prior on $w_{1:K-1}$ is
$
w_{1:K-1}  \sim  \mathcal{D}ir(\delta)
$
where $\mathcal{D}ir(\delta)$ is the symmetric Dirichlet distribution. %Having only the concentration parameter $\delta$ specified means we do not have any prior knowledge favouring one component over another, but we still can control how evenly spread the weights $\bm{w}$ are.

\subsubsection{Distribution of $s^k_{i}$}

Following the hierarchical representation, given $z_i=k$, the distribution of the variance parameter $s^k_i$ in \eqref{condlasso}, is set to be
\begin{equation*}
s_{i}^k \sim  \mathcal{G}a(d/2,d/2)
\end{equation*}
where $\mathcal{G}a(a,b)$ is Gamma distribution of mean $a/b$. The hyperparameter $d$ corresponds to the degrees of freedom of the student-$t$ distribution. %This is obtained by integrating an infinite mixture of normal over a gamma distributed variance parameter. 
%The choice of a lower degrees of freedom parameter $d$ allow us to build a robust regression model that can accommodate for observations errors or more extreme outliers. 

%In Figure \ref{GammaPrior} we can see how the shape of the prior changes as the degrees of freedom increase.

%\begin{figure}[!h]
  % \centerline{\includegraphics[height=10cm, width=14cm, angle=0]{GammaPrior.pdf}}
   %\caption{Gamma prior distribution on the dispersion parameter $s_i^k$. We can see how the shape of the distribution changes as a function of the degrees of freedom of the Student's $t$ regression error. \label{GammaPrior}}
 %\end{figure}

\subsubsection{The Bayesian Lasso}
A very important feature of the model we propose is that it combines, in a mixture framework, shrinkage and variable selection. It achieves this result by adopting specific priors for the regression coefficients $\bm{\beta}$ and  the binary indicator variables $\bm{\gamma}$.
 
Tibshirani (1996)
showed that using a ML approach in a single mixture component framework,  one can regularise the estimated linear regression coefficients $\beta_{1:p}^{k}$ introducing the penalty term: $h_\lambda (\beta_{1:p}^{k})=\sum_{d=1}^{p} |\beta^k_d|^q$ for some $q \geq 0$ and $\lambda_k \in \mathbb{R}^+$. The effect of penalising the likelihood function is to shrink the vector of MLE of $\beta_{1:p}^{k}$ toward zero with the possibility of setting some coefficients exactly equal to zero.

It is well known that similar results to the Lasso penalty can be achieved by assuming that $\beta_{1:p}^k$ have independent Laplace, i.e.~double-exponential priors,
\begin{equation}\label{lapdist}
p(\beta_{1:p}^k| \sigma^2_k) = \prod_{d=1}^{p} \frac{\lambda_k}{2\,\sigma_k}\, \exp \left({\frac{-\lambda_k |\beta_{d}^k|}{\sqrt{\sigma^2_k}}}\right)
\end{equation}
where $\sigma^2_k \in \mathbb{R}^+$ determines the scaling of the regression coefficients in the $k^{th}-$curve and $\lambda_k \in \mathbb{R}^+$ is the smoothness parameter that controls the tail decay. Since the mass of \eqref{lapdist} is quite highly concentrated around zero with a distinct peak at zero, the regression coefficient estimates corresponding to the posterior mean and posterior mode are shrunk towards zero in equivalent fashion to the penalisation least squares estimation procedure. 

The double-exponential distribution can be represented as a scale mixture of normals with exponential mixing distribution. Therefore, introducing a latent vector of scale variables we obtain a more tractable hierarchical formulation of the prior on $\beta^k_{1:p}$. Ignoring for the moment the $\bm{\gamma}_{1:p}$ indicator and assuming a single component mixture, consider the following hierarchical prior on the $d^{th}$ regression coefficient: $\beta_d| \tau^2_d,\lambda \sim \mathcal{N}(0,\tau^2_d)$ where the hyperparameter $\tau^2_d$ itself has hyperprior $\tau^2_d \sim \mathcal{E}x(\lambda^2/2)$ ($\mathcal{E}x(a)$ is the exponential distribution of mean $1/a$.). We note that marginally $\beta_d$ still follows a Laplace distribution with parameter $\lambda$, 
$
p(\beta_d)=\int_{0}^{\infty} \, p(\beta_d |\tau^2_d) \, p(\tau^2_d)\, d\tau^2_d \propto \exp(-\lambda \, |\beta_d|).
$

The modular structure of hierarchical modelling allows us to extend, in a straightforward way, the Bayesian Lasso method to our proposed mixture of linear regression. Together with the prior on $\beta_{\gamma_{1:p}^k}^k$ we also specify priors on the hyperparameters $\sigma_k$, with $\sigma_k \in \mathbb{R}^+$, to control the scaling, and $\tau_{1:p}^{k}$, with $\tau_{d}^{k}\in \mathbb{R}^+ $, to induce shrinkage on the coefficients of the $k^{th}$ regression curve.
\begin{eqnarray*}
\beta_{\gamma_{1:p}^k}^k|\sigma_k^2, \tau_{\gamma_{1:p}^k}^{2,k},\gamma_{1:p}^k & \sim & \mathcal{N}_{|\gamma_{1:p}^k|_1}\bigg(0,\sigma_k^2\textrm{diag}(\tau_{\gamma_{1:p}^k}^{2,k})\bigg)\\
\sigma_k^2 & \sim & \mathcal{IG}a(a,b)\\
\tau_{\gamma_{1:p}^k}^{2,k}| \gamma_{1:p}^k & \stackrel{\textrm{i.i.d.}}{\sim} & \mathcal{E}x(\lambda^2/2).
\end{eqnarray*}
$\mathcal{IG}a(a,b)$ is the Inverse-Gamma distribution of mean $b/(a-1)$ ($a>1$).   Whilst in our discussion we assume $\lambda$ is given, Park \& Casella (2008) have shown, in a non-mixture Bayesian framework with known $\gamma_{1:p}$, that the Lasso parameter can be chosen by marginal maximum likelihood or using an appropriate hyperprior. 

\subsection{Variable Selection}

%The frequentist version of the Lasso provides a straightforward method for variable selection by identifying the non-important predictor variables as those variables whose $\hat{\beta}^k_d = 0$. In the Bayesian version of the Lasso approach, under the absolutely continuous (w.r.t. Lebesgue measure) double-exponential prior distribution, the prior probability of the event $\{\beta^k_d = 0 \}$ is zero. Thus, the posterior probability of such an event must also be zero. To overcome this problem, we are required to explicitly allocate prior probability mass to these events, $\{\beta^k_d = 0 \}$ in order for posterior inferences about events to be coherent.

%Placing prior mass on the event $\{\beta^k_d = 0\}$ is equivalent to assigning a prior distribution to the space of two alternative regression models: one which includes the $d^{th}$ covariate and the other one which excludes it. This can be done in a Bayesian framework using the latent indicator variable, $\gamma^k_{1:p}$, where $p(\gamma^k_d = 1)$ corresponds to prior probability of the model including the variable $x_d$ and $p(\gamma^k_d = 0)$ indicates to probability the alternative event.

Many authors, such as George \& McCulloch (1997),  Kim et al.~(2006)  and  Sch\"afer \& Chopin (2012)  have proposed an effective solution by once again specifying a convenient prior for the selection indicator $\gamma^k_d$. 
We specify selection priors that fit into the mixture framework of regularised regressions. A suitable prior for $\gamma^k_d$ is the Bernoulli distribution $\mathcal{B}e(\phi)$  mutually independent across independent
components.
%, yields
%\begin{equation*}
%p(\gamma^k_{1:p}) = \prod_{d=1}^{p} \phi^{\gamma^k_d}\,(1-\phi)^{1-\gamma^k_d}.
%\end{equation*}
%Note that setting $\phi=1/2$ means that when considering whether to include the variable $x_d$ we do not have any prior information and the two alternative models, $\gamma^k_{d}=1$ and $\gamma^k_{d}=0$, are equally likely. Only after having observed the data, the important predictor variables can then be identified by examining the marginal posterior inclusion probabilities.

We should also point out the level of the flexibility of the mixture model. By making $\gamma^k_{1:p}$ cluster specific, each regression curve can be a function of its own different set of covariates. On the other hand, the combinations of competing models to be evaluated grows exponentially with the number of explanatory variables and  linearly with the clusters, $K2^{p}$. In theory, for the given prior, we could compute the posterior probability of each model before selecting the best one. In practice, it is evident that a full exploratory search is unfeasible and we need to incorporate a selection procedure into the sampling algorithm.

%In section \ref{sec:regmeth} we discuss how to construct a stochastic sampler which will allow us to generate samples from the marginal posterior distribution of $\pi(\gamma^k_{1:p})$. This solution will provide a viable computational approach for addressing model uncertainty while preserving the regularization properties induced by the Bayesian Lasso methodology. We first need to derive the posterior distribution of the parameters of interest.

\subsection{Posterior Distribution}

\begin{figure}[!h] 
   \centerline{\includegraphics[height=10cm, width=14cm, angle=0]{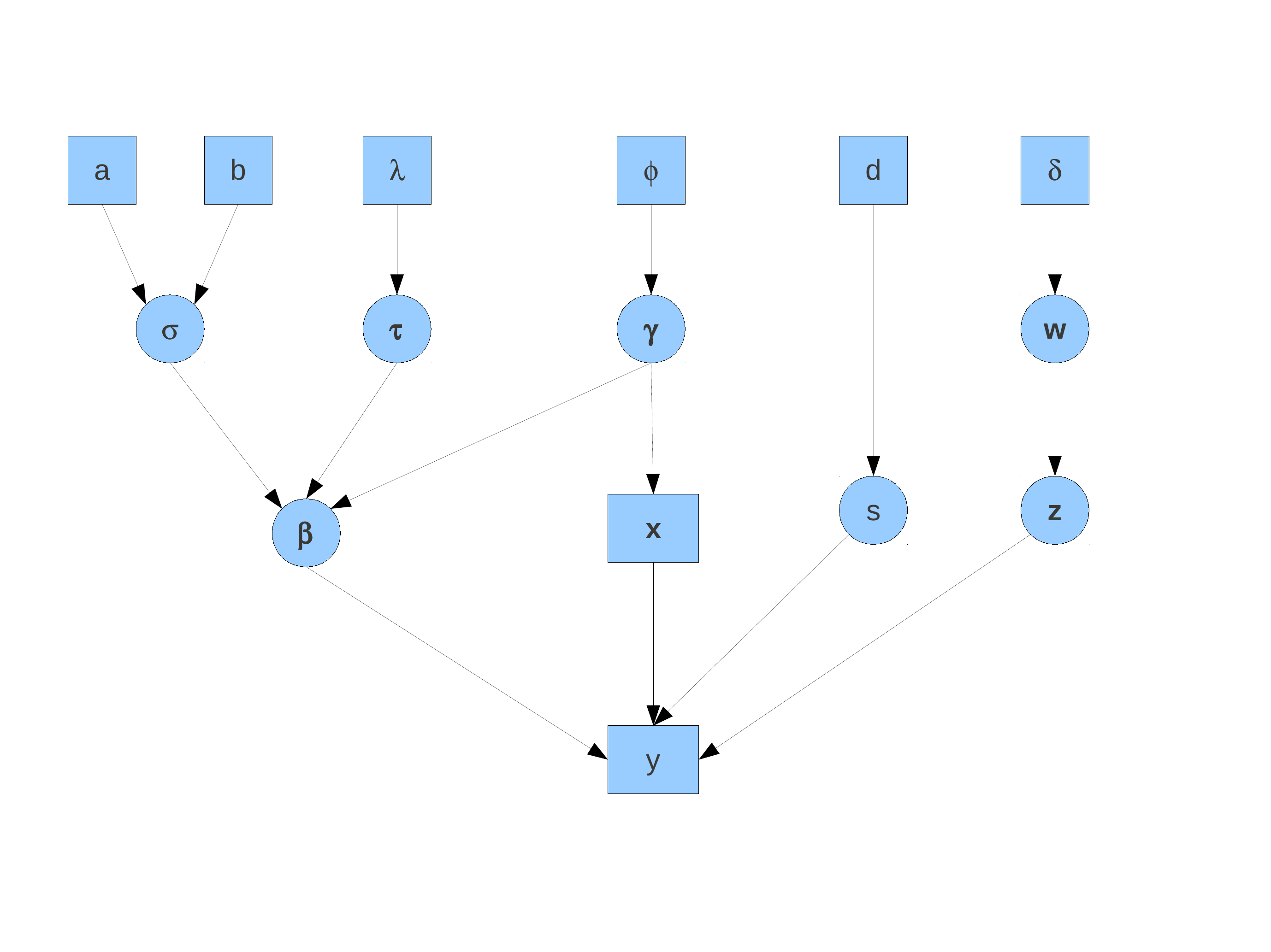}}
   \caption{Directed Acyclic Graph (DAG) showing the hierarchical structure of the priors on the parameters of the proposed mixture model. We have drawn a square box around hyperparameters considered to be a known constant, a circle to indicate an latent variables that need to be estimated, and a rectangular box to indicate observed data. The arrows indicate the conditional dependence structure of the model.\label{fig:dag}}
 \end{figure}

The hierarchical representation of our Bayesian model can be observed in Figure \ref{fig:dag}.
We have also discussed in the previous section how the desired properties of the model are achieved by specifying a convenient structure. We now give some details on the posterior of interest

%----------------
Using a synthetic notation to indicate the unknown parameters of the model $\bm{\psi} = (w_{1:K},\sigma_{1:K}, \bm{\beta}_{1:K}, \bm{s}_{1:n}, \bm{\gamma}_{1:p}, \bm{\tau}^2_{1:p})$, and the fixed, assumed known, hyperparameters of the model $ \bm{h} = (a, b, \lambda, \phi, d, \delta )$, we can say that, after observing the covariates $\bm{x}=(\bm{x}_1,\ldots,\bm{x}_n)$ and the responses $\bm{y}=(y_1,\ldots,y_n)$, the posterior distribution of $\bm{\psi}$ is
 \begin{equation}\label{eq:genpost}
\pi(\bm{\psi}|\bm{x},\bm{y}) \propto L(\bm{y};\bm{x},\,\bm{\psi}) \, p(\bm{\psi}|\bm{h})
\end{equation}
where $ L(\bm{y};\bm{x},\bm{\psi})$ is the likelihood function and $p(\bm{\psi}|\bm{h})$ the prior distributions we have previously defined.

Since our main focus is to draw inference on the cluster membership of the observations and identify the relevant explanatory variables, we remove as many other variables as possible. We integrate out the parameters $\bm{\beta}_{1:K}$, $\sigma_{1:K}$ and $w_{1:K-1}$ in \eqref{eq:genpost}
\begin{equation*}
\pi (z_{1:n},\bm{s}_{1:n},\bm{\gamma}_{1:p},\bm{\tau}^2_{1:p}|\mathcal{D}_n)=  \int \pi(z_{1:n},\bm{s}_{1:n},\bm{\gamma}_{1:p},\bm{\tau}^2_{1:p},\bm{\beta}_{1:K}, \sigma_{1:K},w_{1:K-1}|\mathcal{D}_n ) \, d(\bm{\beta}_{1:K}, \sigma_{1:K},w_{1:K-1} ) 
\end{equation*}
and obtain the marginal posterior density of interest up to a normalizing constant
\begin{align}\label{eq:post}
\pi(z_{1:n},\bm{s}_{1:n},\bm{\gamma}_{1:p},\bm{\tau}^2_{1:p}|\mathcal{D}_n)
& \propto& \prod_{k=1}^K \bigg[\xi_j(s^k_{1:n},\gamma^k_{1:p},\tau_{1:p}^{2,k}|\widetilde{\mathcal{D}}_k) \bigg\{ \prod_{i=1}^n \varphi(s^k_i;d/2,d/2)\bigg\}\times \notag  \\
& & \bigg \{\prod_{d:\gamma_d^k\neq 0} \varphi(\tau_d^{2,k};1,\lambda^2/2) \bigg\}
\bigg]\, \frac{\prod_{k=1}^K\Gamma(\delta+n_k)}{\Gamma(\sum_{k=1}^K[n_k+\delta])}
\end{align}
where $\Gamma(\cdot)$ is the gamma function, $\varphi(x;a,b)$ is the Gamma density of mean $a/b$, $n_k=\sum_{i=1}^n\mathbb{I}_{\{k\}}(z_i)$ the number of observations assigned to the $k^{th}$ cluster, $\widetilde{\mathcal{D}}_k$ is the collection of observations assigned to the $k^{th}$ cluster. Given $z_i=k $, we can derive 
\begin{equation*}\label{eq:xifun}
\xi_k(s_{1:n}^k,\gamma_{1:p}^k,\tau_{1:p}^{2,k}|\widetilde{\mathcal{D}}_k) = \frac{|V_k^*|^{1/2}\Gamma(a_k^*)b^a (b_k^*)^{(-a_k^*)}}{|V_k|^{1/2}\pi^{n_k/2}\Gamma(a)}
\end{equation*}
with
\begin{eqnarray*}
V_k & = & \textrm{diag}(1,\tau_{\gamma_{1:p}^k}^{2,k})  \\
V_k^* & = & \bigg(\textrm{diag}(\tau_{\gamma_{1:p}^k}^{2,k})^{-1} + 
x_{\gamma_{1:p}^k}'\Sigma_{s_{1:n}^k}^{-1}x_{\gamma_{1:p}^k}
\bigg)^{-1}\\
m_k^* & = & V_k^*(x_{\gamma_{1:p}^k}'\Sigma_{s_{1:n}^k}^{-1}y^{k})\\
a_k^* & = & a + n_k/2\\
b_k^* & = & b + \bigg( (y^j)'\Sigma_{s_{1:n}^k}^{-1}y^k - (m_k^*)'(V_k^*)^{-1}m_k^*\bigg)/2
\end{eqnarray*}
where $\Sigma_{s_{1:n}^k}=\textrm{diag}(s_1^k,\dots,s_n^k)$. We should also be aware of label switching problem (e.g.~Jasra et al.~(2005)) which is a common issue when estimating the parameters of a Bayesian mixture model. This is addressed in Section \ref{sec:exres}.

%With conjugate priors the marginal posterior distribution of model parameters $(\bm{\gamma}_{1:p},\bm{\tau}^2_{1:p})$ and allocation variables $z_{1:n}$ are available in closed form. Nonetheless, sampling from the posterior distribution is the only viable approach that enables us to make inference on arbitrary functionals of the unknown variables. Given the high dimensionality of the posterior distribution we will require an efficient simulation methodology.

\section{Simulation Methodology}\label{sec:regmeth}

%The main tool available to Bayesian inference for sampling from a target distribution, such as $\pi$ in \eqref{eq:post}, are Markov chain Monte Carlo methods; see Robert \& Casella (2004) for a complete review. In line with this approach, here we implement some recently proposed algorithms that have been developed to suit scenarios like ours, involving both an high dimensional model and complex patterns of dependence between parameters.

We adopt an MCMC strategy (see Robert \& Casella (2004) for a review) to sample from the target distribution. 
We should first note that, within the mixture modelling literature, there has been work done on perfect sampling and direct sampling, making use of the full conditional distributions. For example, Mukhopadhyay \& Bhattacharya (2011) proposed a perfect sampling methodology for fitting mixture models. Fearnhead \& Meligkotsidou (2007) instead proposed a direct sampling method that returns independent samples from the true posterior. Unfortunately, the described algorithms have limited applicability in our context.

%Considering that our proposed model allows for a random number of covariates, we need an algorithm flexible enough to explore a parameter space whose dimension is itself a random variable. Confronted with similar problem, Sch\"afer \& Chopin (2012) used a Sequential Monte Carlo (SMC) algorithm to adaptively sample from a binary distribution. Using the variable selection problem in linear regression as test case, they showed that even in difficult circumstances, with hundred of covariates, the Sequential Monte Carlo method can outperform standard techniques based on simple Markov chain exploration.

In light of recent work presented by  Andrieu et al.~(2010), we adopt a Particle Markov chain Monte Carlo (PMCMC) simulation procedure which combines MCMC and SMC methods and takes advantage of the strengths of both. The key feature of PMCMC algorithms is that they are in fact exact approximations of idealised MCMC algorithms, while they use sequential Monte Carlo methods to build high dimensional proposal distributions. On the other hand, compared to stand alone SMC, PMCMC sampling is more robust to the path degeneracy problem, described later on. More precisely, here we implement a particle Metropolis-within-Gibbs algorithm. Below we describe the constituents of the algorithm, which is summarized in Section \ref{sec:sampproc}.
% which is effective in situations where using the prior distribution of the underlying latent process as the proposal distribution is the only known practical solution.

%It is worth pointing out that, as we iterate through the simulation algorithm, the cluster structure evolves with the choice of variables and we should appreciate the fact that the variable selection, in the context of clustering, is much more complicated than in the standard classification or %regression analysis. 

%----------------------------------------------------------------
\subsection{Sequential Monte Carlo Algorithm}\label{sec:smcalgo}

%--- Intro ---
SMC methods are a general class of algorithms that use a set of weighted particles to recursively approximate a sequence of distributions of increasing dimension. It has been originally introduced to deal with situations with dynamic observations. Nonetheless, it has demonstrated to be highly effective also in static problems like mixture models and it is an integral part of PMCMC.
Before illustrating how SMC algorithms are used in our sampling procedure, we refer the reader to Doucet et al.~(2001)  for a detailed review of SMC methods. In particular, we assume the reader is familiar with Sequential Importance Sampling (SIS). 

\subsubsection{Sampling Cluster Labels}

%The SMC method allows us to approximate the conditional posterior distribution of the latent label indicator variables $\pi_n(z_{1:n} | \bm{s}_{1:n}, \bm{\gamma}_{1:p}, \bm{\tau}^2_{1:p}, \mathcal{D}_n)$. Following Algorithm \ref{al:smc}, we first initialize $\bm{s}_{1:n},\bm{\gamma}_{1:p},\bm{\tau}^2_{1:p}$ by sampling their respective priors, and then alternate sequential importance sampling and resampling steps. 

We use an SMC method to sample sequentially from $\pi_i(z_{1:i}|\bm{s}_{1:i},\bm{\gamma}_{1:p},\bm{\tau}^2_{1:p},\mathcal{D}_i)$ as $i$ increases.
Following Algorithm \ref{al:smc}, we first initialize $\bm{s}_{1:n},\bm{\gamma}_{1:p},\bm{\tau}^2_{1:p}$ by sampling their respective priors, and then alternate sequential importance sampling and resampling steps. 
More explicitly, the sequential importance sampling targets the full conditional density of the latent labels variables $z_{1:i}$ which, after the first $1,\ldots,i$ data points, is
\begin{equation*}
\pi_i(z_{1:i}|\bm{s}_{1:i},\bm{\gamma}_{1:p},\bm{\tau}^2_{1:p},\mathcal{D}_i)
\propto \bigg[\prod_{k=1}^K \xi_k(s_{1:i}^k,\gamma_{1:p}^k,\tau_{1:p}^{2,k}|\widetilde{\mathcal{D}}_k^{(i)}) \Gamma(\delta + n_j^{(i)})\bigg]\bigg/\Gamma(\sum_{k=1}^K(n_k^{(i)}+\delta))
\end{equation*}
where $\widetilde{\mathcal{D}}_k^{(i)}$ denotes the  data allocated to the $k^{th}$ cluster out of the first $i$ observations and 
$n_k^{(i)}=\sum_{l=1}^i\mathbb{I}_{\{k\}}(z_l)$
their total number.

%--------------------------------------------
\subsubsection{Adaptive Resampling}\label{subsec:resamp}

SIS is subject to the problem of weight degeneracy. As new incoming observations are fed into the algorithm, the variance of the importance weights typically increases at an exponential rate until all the mass concentrates on one single particle, leaving the remaining particles with weights tending to zero.

To avoid spending a large computational effort to update trajectories whose contribution to the final estimate is negligible, we execute a resampling step with the intention of replacing the unpromising lowest weighted particles with new particles that hopefully lie in regions of high target density. The exact procedure consists in sampling $N$ particles from the approximated target distribution to obtain $N$ new particles which will then be equally weighted. 
On the other hand, if one resamples very often, we will rapidly deplete the number of distinct particles and the approximation of the target will suffer because the paths of $z_{1:i}$ become very similar (path degeneracy). 

To find a balance between weights degeneracy and path degeneracy,  Del Moral et al.~(2012) among others, suggest to resample only when the variance of the unnormalized weights is above a fixed threshold. In the solution we adopt, the threshold is a function of the Effective Sample Size (ESS)
$
ESS=\left( \sum_{j=1}^N (W_n^j)^2  \right)^{-1}
$
which takes values between 1 and $N$ and, as described in Algorithm \ref{al:smc}, we resample only when it is below $ESS < N/2$. 
%To fully appreciate the effect of introducing this rule, in the experimental section \ref{sec:degen} we tested and compared the two versions of the algorithm, with and without adaptive resampling, and show the different impact they have on weights dispersion and paths diversity. 

It should be noted here that executing the resampling step only when the condition $ESS < N/2$ is satisfied, does not alter the property of the algorithm that still returns an unbiased estimate of the normalising constant, as noted in a personal communication by C. Andrieu  and N. Whiteley -
see the work of Arnaud \& Le Gland (2009).

\begin{algorithm}
\caption{Sequential Monte Carlo Algorithm}
\label{al:smc}
{\bf Step 1.} Sample $N$ labels, $z_1^1,\dots,z_1^N$, from $\pi_1(z_1|\cdots)$ and set the corresponding weights $W_1^j=1$ for $j=1,\dots,N$. \\
{\bf Step 2.} For $i=2,\dots,n$ repeat the following
 \begin{enumerate}
 \item{} If $ESS < N/2$,  for each $j = \{1, \ldots, N \}$ resample $a_{i-1}^j\in\{1,\dots,N\}$ using the discrete distribution
 \begin{equation*}
 \overline{W}_{i-1}^j = \frac{W_{i-1}^j}{\sum_{g=1}^N W_{i-1}^g}.
 \end{equation*}
Otherwise keep all the current particles by $a_{i-1}^j=j$ for $j \in \{1, \ldots, N \}$.

 \item{} Sample, for each $j\in\{1,\dots,N\}$, a label $z_i^j$ from $\pi_i(z_i|\cdots)$ where
 \begin{equation*}
 \pi_i(z_i|\cdots) = \frac{\xi_{z_i}(s_{1:i}^{z_i},\gamma_{1:p}^{z_i},\tau_{1:p}^{2,z_i}|\widetilde{\mathcal{D}}_{z_i})
 \Gamma(\delta + 1 + n_{z_{i}}^{(i-1),a_{i-1}^j})
 }{
 \sum_{z_i=1}^K
 \xi_{z_i}(s_{1:i}^{z_i},\gamma_{1:p}^{z_i},\tau_{1:p}^{2,z_i}|\widetilde{\mathcal{D}}_{z_i})
 \Gamma(\delta + 1 + n_{z_{i}}^{(i-1),a_{i-1}^j})
 }
 \end{equation*}
 and $n_{k}^{(i-1),a_{i-1}^j}=\sum_{i=1}^{i-1}\mathbb{I}_{\{k\}}(z_i^{(a_{i-1}^{j})})$. Set $z_{1:i}^j=(z_{1:i-1}^{a_{i-1}^j},z_i^j)$.

 \item{} Set, for each $j\in\{1,\dots,N\}$
 \begin{equation*}
 W_i^j = \frac{\sum_{z_i=1}^K\bigg[\prod_{k=1}^K \xi_k(s_{1:i}^k,\gamma_{1:p}^k,\tau_{1:p}^{2,k}|\widetilde{\mathcal{D}}_k^{(i)})
 \Gamma(\delta + n_k^{(i),j})\bigg]\bigg/\Gamma(\sum_{k=1}^K(n_k^{(i),j}+\delta))}
 {\bigg[\prod_{k=1}^K \xi_k(z_{1:i-1}^k,\gamma_{1:p}^k,\tau_{1:p}^{2,k}|\widetilde{\mathcal{D}}_k^{(i-1)})
 \Gamma(\delta + n_k^{(i-1),j})\bigg]\bigg/\Gamma(\sum_{k=1}^K(n_k^{(i-1),j}+\delta))}
 \end{equation*}
 and $i=i+1$.
 
 \end{enumerate}

\end{algorithm}

\subsection{Conditional Sequential Monte Carlo Algorithm} \label{sec:condsmcalgo}

The conditional SMC algorithm we iterate in the second stage of our sampling procedure is essentially the SMC algorithm described in Section \ref{sec:smcalgo} except it preserves the path of one particle. 

To describe the algorithm, we need to introduce a sequence of indexes $b_{1:n}^{t} \in \{1,\ldots,N\}^n$ to represent the genealogy of the $t^{th}$ particle for $t\in \{1,\ldots,N\}$. Once we have set $b_n^t=t$, the genealogy of $t^{th}$ particle can then be defined recursively $b_i^t=a_{i-1}^{b_i^t}$ for $i=1,\ldots,n-1$ where the $\bm{a}_{1:n-1}= (a^1_{1:n-1},\ldots,a^N_{1:n-1})$ are the recorded samples from the previous iteration of the SMC algorithm.

As we can see from the Algorithm \ref{al:consmc}, the sampling sequence is similar to what is implemented in a standard SMC algorithm except that one randomly chosen particle $t$ with its ancestral lineage $b_{1:n}^{t}$ is fixed and ensured to survive, whereas the remaining $N-1$ particles are regenerated as usual.

\begin{algorithm}

\caption{Conditional Sequential Monte Carlo Algorithm}
\label{al:consmc}

{\bf Step 1.} Sample $1-N$ labels $z_1^j$ from $\pi_1(z_1|\cdots)$, for $j=1,\ldots, N$ while $j \neq b_1^t$ (i.e. excluding $j=b_1^t$), and set all the weights $W_1^j=1$ for $j=1,\dots,N$. \\
{\bf Step 2.} For $i=2,\dots,n$ repeat the following
 
\begin{enumerate}

 \item{} If $ESS < N/2$, for each $j \in \{1, \ldots, N \}$ except $j= b_i^t$, resample $a_{i-1}^j\in\{1,\dots,N\}$ using the discrete distribution
 \begin{equation*}
 \overline{W}_{i-1}^j = \frac{W_{i-1}^j}{\sum_{g=1}^N W_{i-1}^g}.
 \end{equation*}
  Otherwise keep all the current particles by $a_{i-1}^j=j$.

 \item{} Sample $z_1^j$ from $\pi_i(z_i|\cdots)$ for each $j \in \{1, \ldots, N \}$ except $j= b_i^t$, and update the corresponding path $z_{1:i}^j=(z_{1:i-1}^{a_{l-1}^j},z_i^j)$.

 \item{} Set $W_i^j$ as for the SMC algorithm, for each $j\in\{1,\dots,N\}$, (this includes the fixed particle $j=b_i^t$)
 \end{enumerate}
\end{algorithm}

\subsection{Markov Chain Monte Carlo Steps} \label{sec:mcmcalgo}

In the SMC algorithm we sample from the posterior distribution of the latent label indicator variable $z_{1:n}$. With the MCMC steps our objective is to update the other parameters of the mixture model that control the regression error distribution, the regularization of the regression coefficients and the variable selection process. 
The MCMC steps which target the posterior \eqref{eq:post} are as follows. 

\subsubsection{Step 1: Update $\bm{\tau}_{1:p}^2$}

To update the $\bm{\tau}_{1:p}^2$, given all the other variables are fixed, we can use the following procedure. For each $k\in\{1,\dots,K\}$, assuming $|\gamma_{1:p}^k|_1>0$, sample for each $d$ where $\gamma_d^k=1$, 
$
(\tau_{d}^{2,k})^* = \tau_{d}^{2,k}\exp\{\nu_{\tau}\, N_d\}$
with $\nu_{\tau}>0$ a user-set parameter and $N_d\sim\mathcal{N}(0,1)$, independent for each $d$. Accept all the $(\tau_{d}^{2,k})^*$ with probability
\begin{equation*}
1\wedge\frac{\xi_{k}(s_{1:n}^k,\gamma_{1:p}^k,(\tau_{1:p}^{2,k})^*|\widetilde{\mathcal{D}}_k)}
{\xi_{k}(s_{1:n}^k,\gamma_{1:p}^k,\tau_{1:p}^{2,k}|\widetilde{\mathcal{D}}_k)}\prod_{d;\gamma_{d}^k\neq 0}\frac{\varphi((\tau_{d}^{2,k})^*;1,\lambda^2/2)(\tau_{d}^{2,k})^*}
{\varphi(\tau_{d}^{2,k};1,\lambda^2/2)\tau_{d}^{2,k}}
\end{equation*}
otherwise keep the current $\tau_{1:p}^{2,k}$.

\subsubsection{Step 2: Update $\bm{s}_{1:n}$}

To update $\bm{s}_{1:n}$, given all the other variables are fixed, we can use the following procedure. For each $i\in\{1,\dots,n\}$, $k\in\{1,\dots,K\}$ propose
%\begin{equation}\label{eq:updates}
$
(s_{i}^{k})^* = s_{i}^{k}\exp\{\nu_s N_i\}
$
%\end{equation}
where $\nu_s>0$ is a user-set parameter (potentially different from the $\nu_{\tau}$ above) and $N_i\sim\mathcal{N}(0,1)$, independent for each $i$. Note that $(s_{1:n}^k)^*$ features only one changed value from $s_{1:n}^k$. The proposed move then is accepted with probability
\begin{equation*}
1\wedge \frac{\xi_{k}((s_{1:n}^k)^*,\gamma_{1:p}^k,\tau_{1:p}^{2,k}|\widetilde{\mathcal{D}}_k)}
{\xi_{k}(s_{1:n}^k,\gamma_{1:p}^k,\tau_{1:p}^{2,k}|\widetilde{\mathcal{D}}_k)}\prod_{i;\gamma_{i}^k\neq 0}\frac{\varphi((s_{i}^{k})^*;d/2,d/2)(s_{i}^{k})^*}
{\varphi(s_{i}^{k};d/2,d/2)s_{i}^{k}}
\end{equation*}
otherwise keep the current $s_{i}^{k}$. 

\subsubsection{Step 3: Update $\bm{\gamma}_{1:p}$}

To update $\bm{\gamma}_{1:p}$, given all the other variables are fixed, we can use the following procedure. For each $d\in\{1,\dots,p\}$, $k\in\{1,\dots,K\}$ (i.e.~propose to change only one element each time), if $\gamma_d^k=0$ we propose $(\gamma_d^k)^*=1$ and draw $(\tau_d^k)^*$ from its prior ($\mathcal{G}a(1,\lambda^2/2)$).
The proposed move is accepted with probability
\begin{equation*}
1 \wedge\frac{\xi_{k}(s_{1:n}^k,(\gamma_{1:p}^k)^*,(\tau_{1:p}^{2,k})^*|\widetilde{\mathcal{D}}_k)}
{\xi_{k}(s_{1:n}^k,\gamma_{1:p}^k,\tau_{1:p}^{2,k}|\widetilde{\mathcal{D}}_k)}
\end{equation*}
otherwise we keep $\gamma_d^k=0$. If $\gamma_d^k=1$, we propose to set it to be zero, removing the corresponding $\tau_d^{2,k}$ and using the same expression as above to accept/reject (with the appropriate changes i.e. the proposed state here has fewer variables than the current model). In this proposal, we are adding or removing columns from our design matrix.

Note that this algorithm is best suited for scenarios similar to the ones we investigate in this paper, where the number of components $K\geq 2$ and the number of data points $n \geq 30$ make the space to be sampled much bigger than the one for the explanatory variables.

\subsection{Sampling Procedure}\label{sec:sampproc}

%We can effectively achieve the result of simulating from the posterior distribution \eqref{eq:post}, following a two stage procedure. In the preliminary stage, we initialise the algorithm by sampling plausible parameters values from the corresponding priors and generate various particles that represent possible cluster assignments of the observed data points. In the subsequent stage, which should be repeated until convergence, we alternate a conditional SMC step, which produce a likely labelling of the data, and a Metropolis step, which updates the error estimate and the other parameters of interest.

The sampling procedure consists of

\begin{itemize}
\item{\bf {Stage I:}} Initialise the algorithm. Sample $\bm{s}_{1:n},\bm{\gamma}_{1:p},\bm{\tau}^2_{1:p}$ from the respective priors. Run the SMC algorithm, as described in section \ref{sec:smcalgo}, storing all the $N$ particles labels $\bm{z}_{1:n}=z_{1:n}^1,\dots,z_{1:n}^N$ and their genealogy $\bm{a}_{1:n-1}=(a_{1:n-1}^1,\dots,a_{1:n-1}^N)$. Sample one particle index $t \in\{1,\dots,N\}$ according to the normalized weights $\overline{W}_n^1, \ldots, \overline{W}_n^N $.
\item{\bf {Stage II:}} Repeat the following steps until convergence
\begin{enumerate}
\item{Run the conditional SMC algorithm, as described in section \ref{sec:condsmcalgo}.}
\item{Sample  $t \in\{1,\dots,N\}$ according to new weights $\overline{W}_n^1, \ldots, \overline{W}_n^N $. Store $z_{1:n}^t$ and $b_{1:n}^t$.}
\item{Given $z_{1:n}^t$, update the current values of  $\bm{s}_{1:n},\bm{\gamma}_{1:p},\bm{\tau}^2_{1:p}$ following the MCMC steps described in section \ref{sec:mcmcalgo}.}

\end{enumerate}

\end{itemize}
This provides a valid MCMC algorithm with the posterior of interest as an appropriate marginal;
see Andrieu et al.~(2010).

%\section{Examples}\label{sec:exres}

%We now present simulation results from simulated and real data.

%The scope of this section is to verify, using simulated datasets, the properties of the mixture model we propose. Testing it in a controlled environment, we aim to highlight the main properties and possible faults of the model.
%
%First we assess how the simulation procedure responds to different scenarios, its sensitivity to changes in prior hyperparameters and how to set the parameters that control the algorithm. In particular, we monitor the acceptance rate of the metropolised Gibbs sampling steps and the degeneracy of the weights and path diversity in the two cases: with resampling that is executed at every iteration and with adaptive resampling which is subject to ESS criterion being satisfied.
%
%The performance of the proposed model is then discussed in terms of clustering accuracy, which means checking that homogeneous observations are correctly grouped together. At the same time we also interested to verify that only the truly informative variables are actually included in the model.
%
%The model simulation procedure and its related sampling algorithms have been coded in Matlab; the code is available on request.
%
%-------------------------------

\section{Simulation Study}\label{sec:exres}

\subsection{Simulation Settings}\label{sec:simset}

We assume one basic scenario that we then perturb to highlight the different properties of the model and different important aspects of the simulation procedure. In the standard scenario the parameters of the model have been randomly generated from the following priors   
$
w_{1:K-1} \sim  \mathcal{D}ir(2)~, 
s_{i}^k \sim  \mathcal{G}a(2,2)~,
\gamma^k_{1:p}\stackrel{\textrm{i.i.d.}}{\sim} \mathcal{B}e(1/2)~,
\tau_{\gamma_{1:p}^k}^{2,k}| \gamma_{1:p}^k \stackrel{\textrm{i.i.d.}}{\sim}  \mathcal{E}x(1/2)~,
\sigma_k^2  \sim  \mathcal{IG}a(2,4)~,
\beta_{\gamma_{1:p}^k}^k|\sigma_k^2, \tau_{\gamma_{1:p}^k}^{2,k},\gamma_{1:p}^k  \sim \mathcal{N}_{|\gamma_{1:p}^k|_1}\bigg(0,\sigma_k^2\textrm{diag}(\tau_{\gamma_{1:p}^k}^{2,k})\bigg)
$
and each data point is then sampled from the mixture model.
%\begin{equation*}
%y_i \, \sim \sum_{k=1}^{K} w_k \, \mathcal{N}( x'_{\gamma_{1:p}^k,i} \, \beta_{\gamma_{1:p}^k}^k , s_i^k).
%\end{equation*}
Each dataset we generate contains $n=50$ paired observations sampled from a mixture of three components $K=3$. The covariates $\bm{x}_i$ of dimension $p=20$ are sampled from a centered Gaussian distribution whose dispersion depends on the cluster membership.

The only parameters of the simulation algorithm we need to set are: the number of particles, say $N=100$; the step length of the MCMC move for $\tau$, say $\nu_{\tau}=2$; the step length of the error update, say $\nu_{z}=3$, and also the number of repeats of the sampling procedure, say a few thousand.

\subsection{Resampling} \label{sec:degen}

An important aspect of the simulation behaviour that we can partially control is the weight degeneracy. By introducing the adaptive resampling step we limit the risk of the empirical probability mass collapsing on a single particle. We are equally aware that resampling tends to replicate the most likely paths and might lead to an impoverished diversity of explored paths. This effect is marginally alleviated by limiting the frequency of the resampling. 

\begin{figure}[!h]
   % \centering
   \begin{minipage}[b]{7cm}
     \includegraphics[angle=0,height=6cm,width=7cm]  
                     {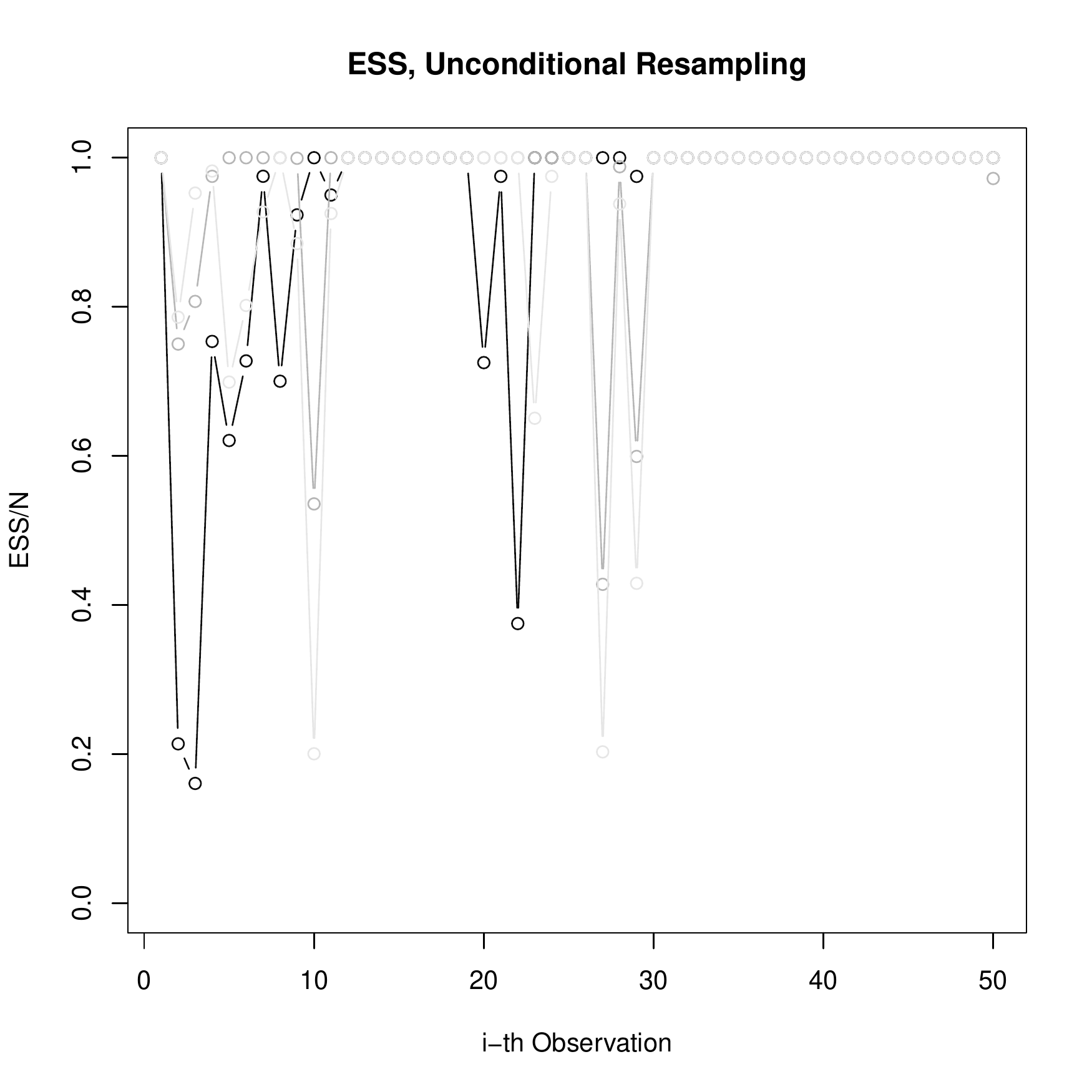}  
   \end{minipage}
   \begin{minipage}[b]{7cm}
     \includegraphics[angle=0,height=6cm,width=7cm]  
                     {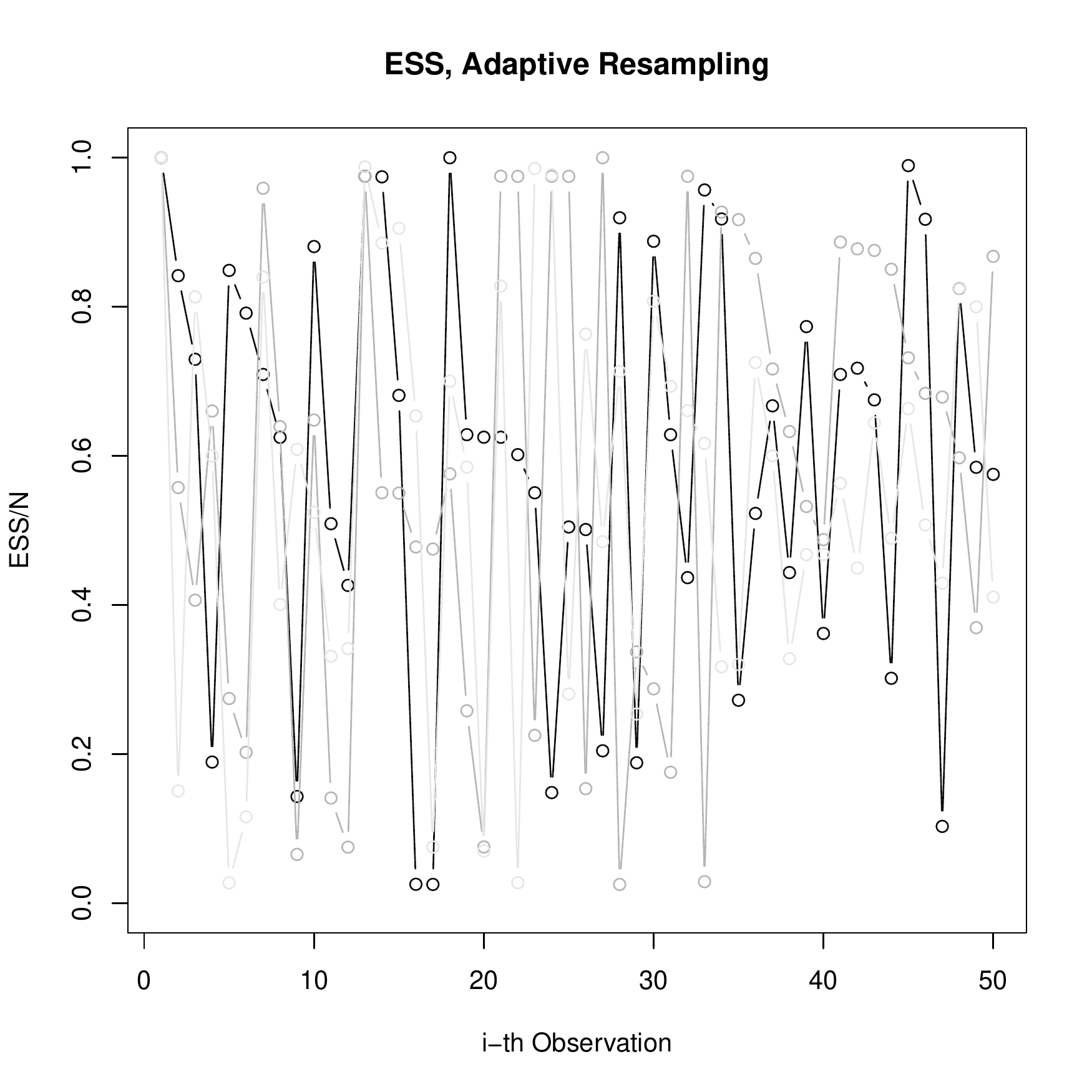} 
   \end{minipage}
   \\[0mm]
   \begin{minipage}[b]{7cm}
     \includegraphics[angle=0,height=6cm,width=7cm]  
                     {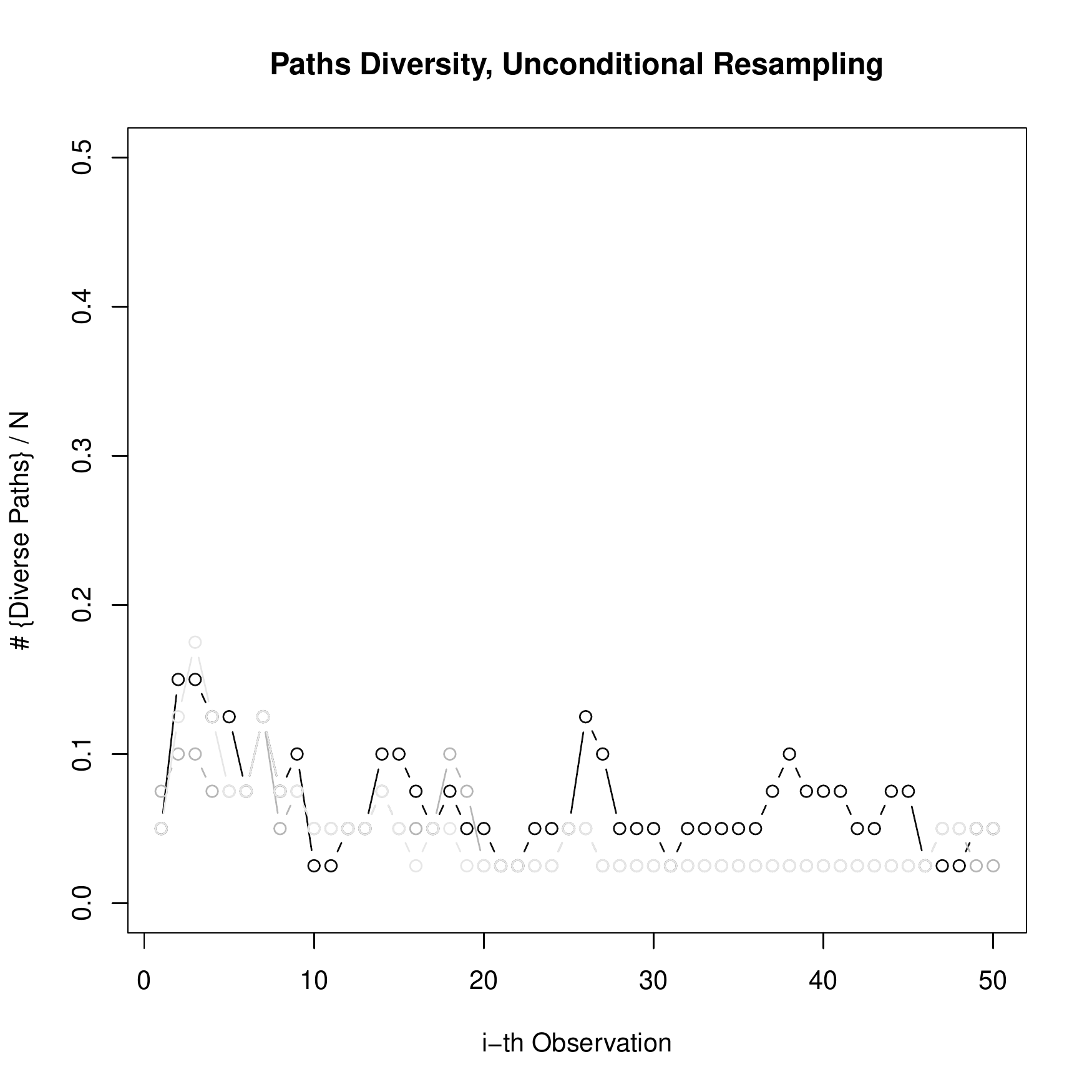}  
   \end{minipage}
   \begin{minipage}[b]{7cm}
     \includegraphics[angle=0,height=6cm,width=7cm]  
                     {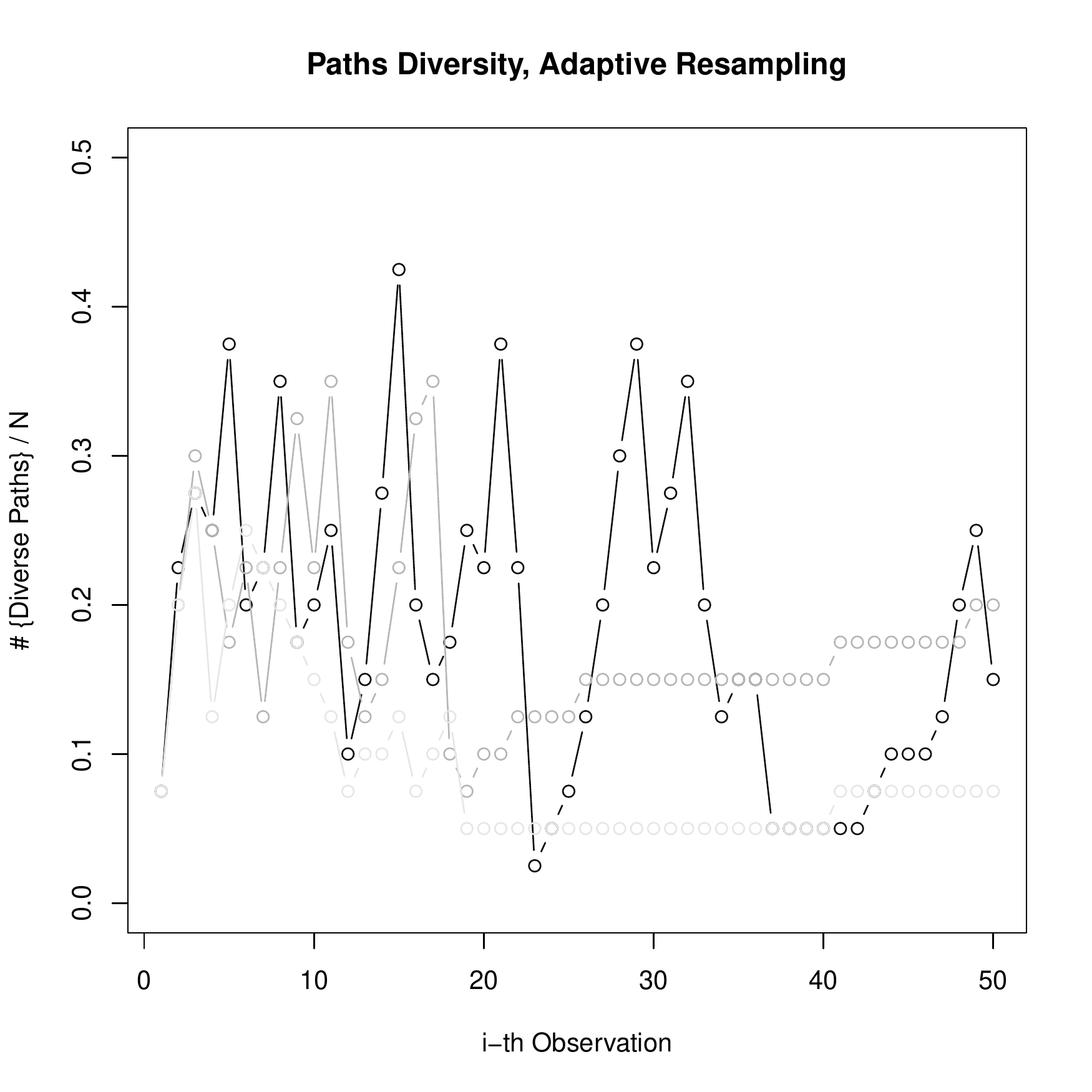} 
   \end{minipage}
   \\[0mm]

  \caption{ \footnotesize{{\bf Left Column:} Unconditional Resampling, we resample systematically every time a new observation is fed into the SMC algorithm. {\bf Right Column:} Adaptive Resampling, we only resample whenever the $ESS$ falls below a fixed threshold. {\bf Top Row:} Weight Degeneracy, measured as $ESS/N$, where $1$ means all particles have equal weight, and $0$ means the entire probability mass is on one particle. {\bf Bottom Row:} Path Degeneracy, measured as percentage of paths that remain different as we loop through the observations. Each line represents three separate repeats of the sampling procedure and darker lines correspond to earlier iterations. }
  \label{fig:weightsdegen}}
 \end{figure}

Figure \ref{fig:weightsdegen} shows that, in our case, adaptive resampling ultimately is beneficial to preserve both path and weight diversity. We note in the left column that if we resample after every new observation is processed, we end up fairly quickly with a single path that gets replicated for all $N$ particle. 
With adaptive resampling, on the other hand, the degeneracy of weights and paths is maintained at a tolerable level. In the right column, we preserve a variety of paths that might have different likelihood as shown by the more disperse ESS plot. Note also how in some instances no resample is performed for several runs and the number of particles remains stable as it is their weight. Even if at the end of every iteration of the Markov chain we only need to store one single particle, it is important that we are able to preserve a richer variety of paths and consequently a more disperse weight distribution from which we can sample.

%----------------------------------
\subsection{Clustering Accuracy}

To test the clustering accuracy of the model, we generate datasets using the simulation settings described in Section \ref{sec:simset}. We then let the algorithm run and for each iteration we save one particle that represents one sample from the posterior distribution of the label indicator variables. Once we have collected enough samples we analyse the distribution of the Adjusted Rand Index Score over the sampled paths. To deal with the label switching problem we permute the labelling to maximize the adjusted rand index, computed w.r.t.~the cluster assignment 
associated to an external model or corresponding to the null hypothesis we want to test (like the macro sector partition in our real life problem).

\begin{figure}[!h]
   \centerline{\includegraphics[height=5cm, width=12cm, angle=0]{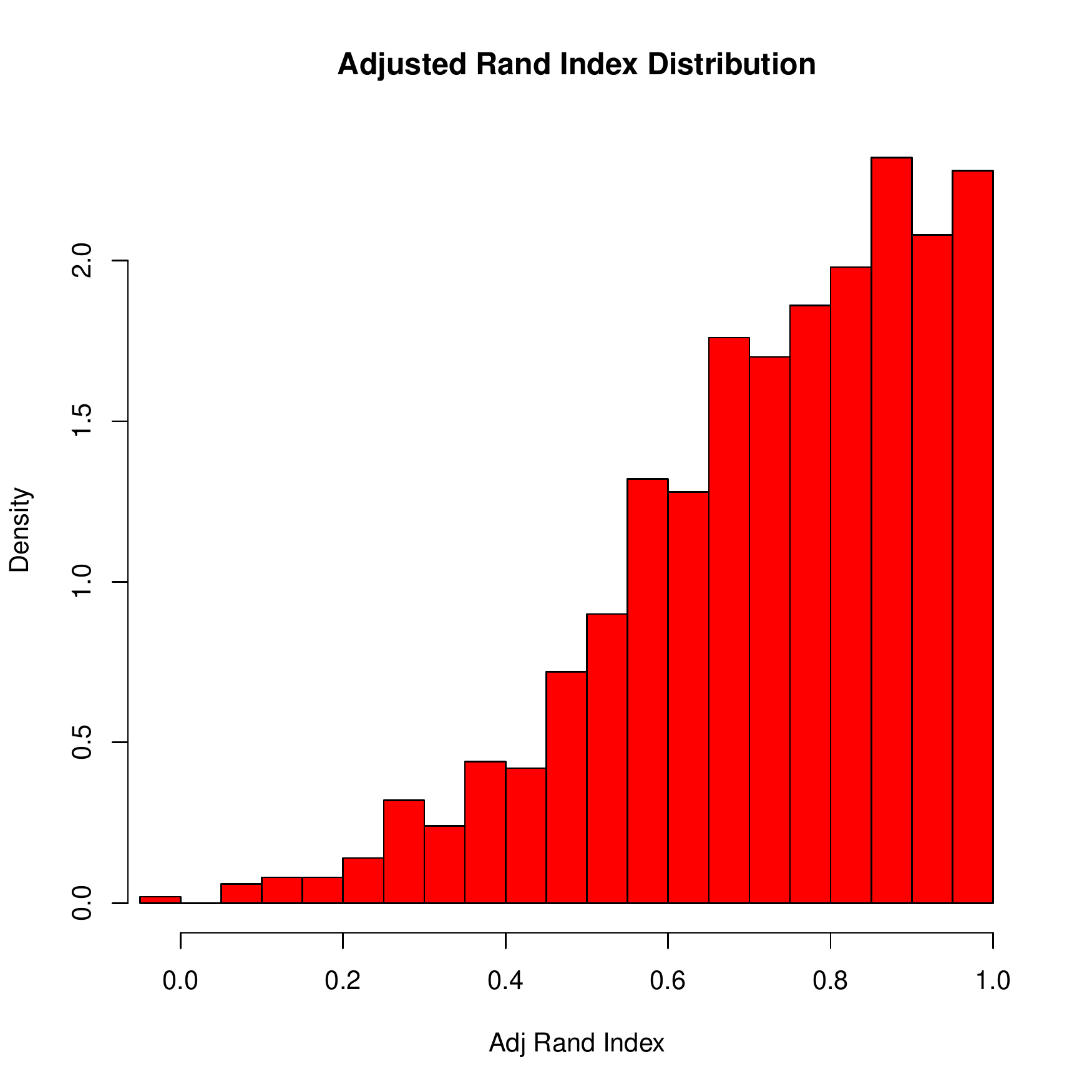}}
   \caption{Adjusted Rand Index distribution. For every MC iteration we record the Adjusted Rand Index score of the proposed cluster assignment versus the true clusters labels. Where a distribution centered around zero would be an indication of random assignment, the observed values give evidence that the model is successfully assigning most of the data points to the proper cluster.\label{fig:adjridx}}

 \end{figure}

In Figure \ref{fig:adjridx} we can see that the distribution is highly skewed towards $1$, which means that most of the time the suggested clustering assignment perfectly matches the true clustering. In other words, given that the classification probability distribution we try to approximate  is fairly accurate (which seemed to be the case on the basis of our convergence assessment) the model seems to provide a good clustering, at least in this example.   

%------------------------------------------
\subsection{Variable Selection Accuracy}

The other major point we want to investigate is the accuracy of the variable selection approach. 
We would hope that the model identifies as many informative variables as possible, and at the same time is sufficiently parsimonious to exclude as many as possible of the noise variables. 
To that end the sensitivity index is the ratio of the number of true variables detected to the sum of the same value added to false negatives and 
specificity index is the ratio of the true negatives to the the sum of the same value added to false positives. These are the measures used to assess the accuracy of the variable selection.

\begin{figure}[!h]
   % \centering
   \begin{minipage}[b]{7cm}
     \includegraphics[angle=0,height=5cm,width=7cm]  
                     {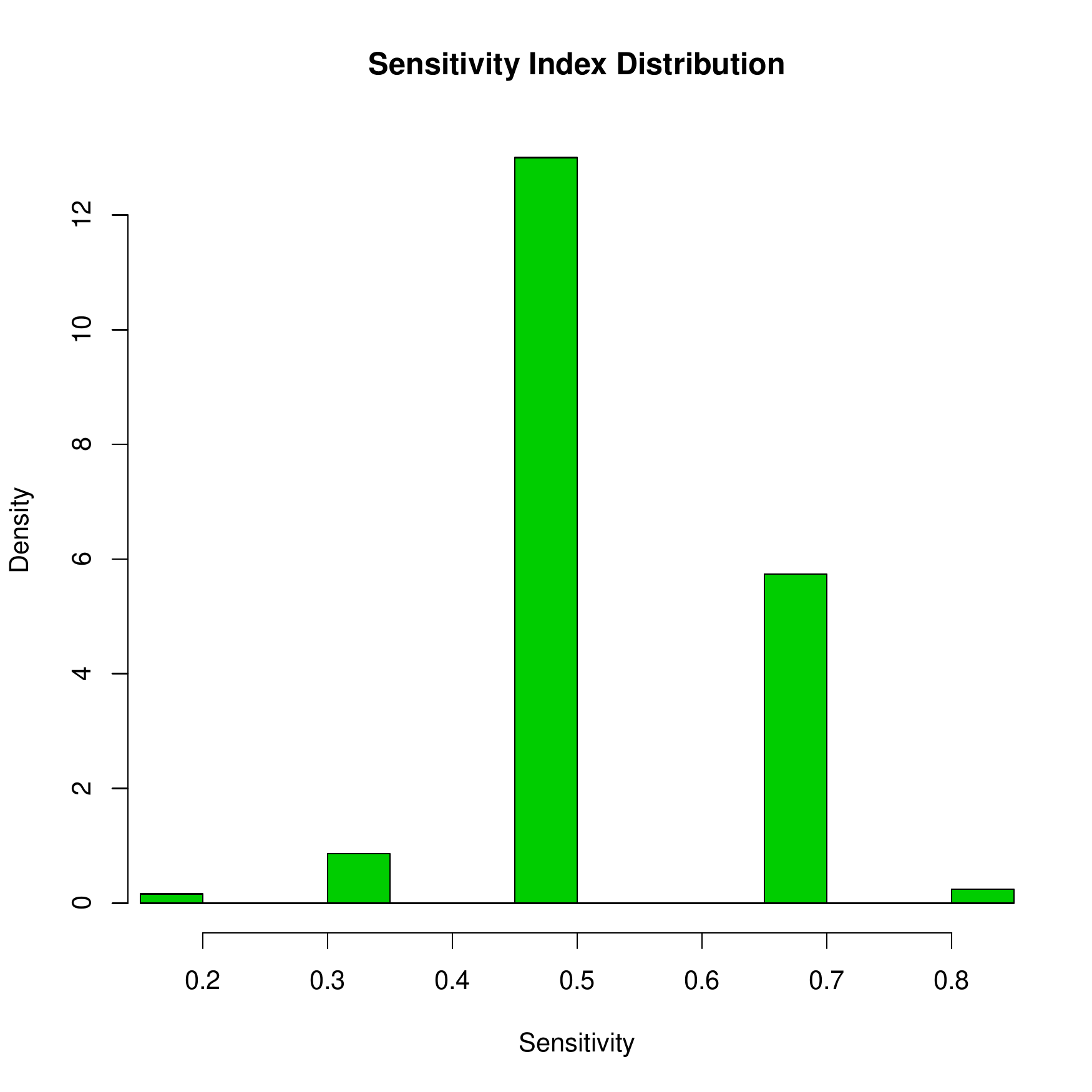}  
   \end{minipage}
   \begin{minipage}[b]{7cm}
     \includegraphics[angle=0,height=5cm,width=7cm]  
                     {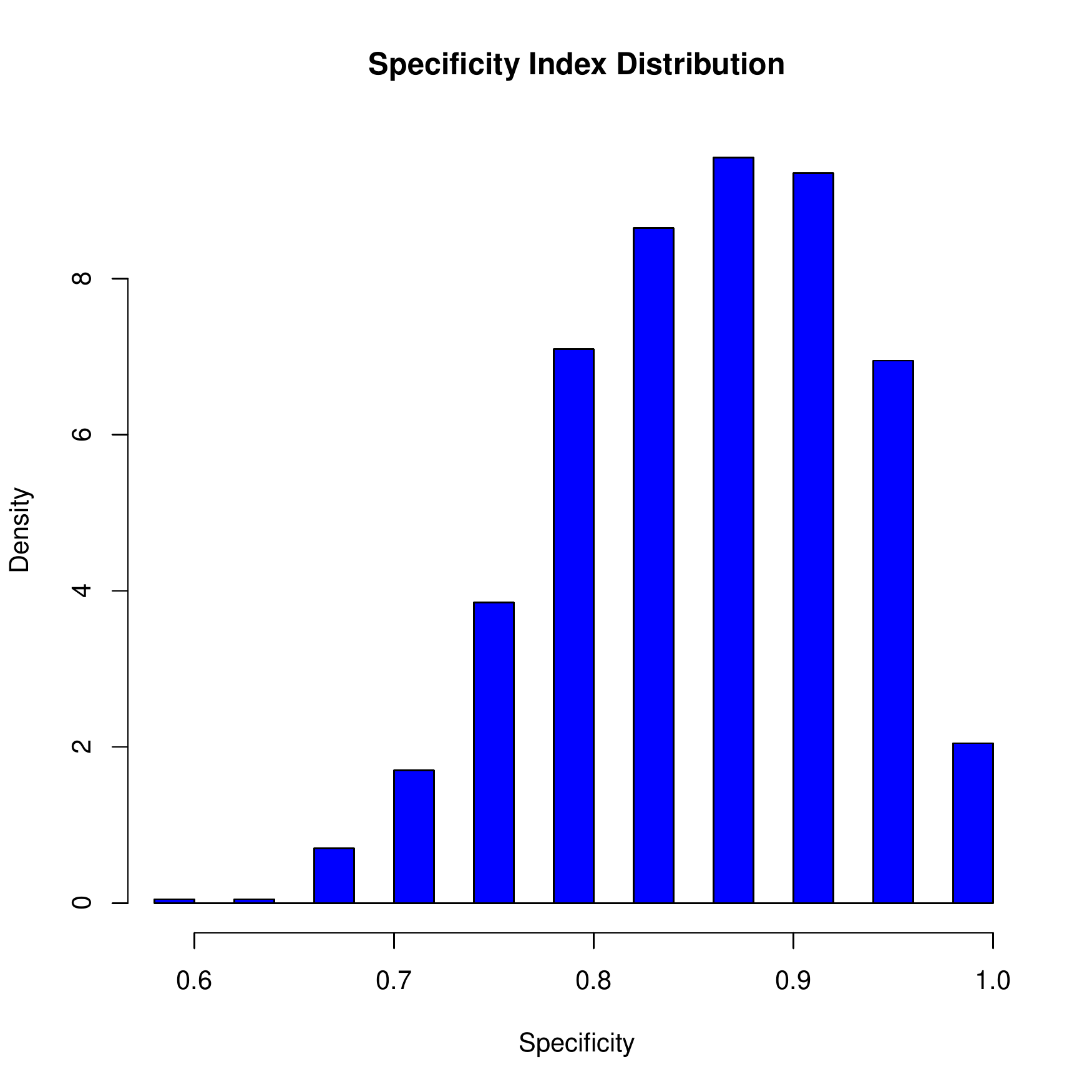} 
   \end{minipage}
   \\[0mm]
  \caption{Variable Selection accuracy over all MC iterations. In the left plot we show the distribution of the sensitivity index, i.e. the ability of the algorithm to identify the truly informative variables. In the right plot the specificity index measures the accuracy of the model in isolating the non-informative variables, which shows that the model is very precise in excluding the noise variables.
  \label{fig:senspec}}
 \end{figure}

In Figures \ref{fig:senspec} we look, as before, at the distribution over all MCMC iterations of the relevant indexes, in this case the sensitivity and specificity indexes. We remark that given the relatively small number of variables, $p=20$, we should not be surprised to observe some very coarse distributions, since there are only so many informative or noise variables. In both plots it is evident that the overall variable selection accuracy is considerable. The sensitivity of the selection algorithm is fairly high, since most of the informative covariates are included and play a role in the regression curves. Conversely, the specificity index is equally good if not better, as very few noise variables are retained at all. We can explain the marginally lower sensitivity compared to the specificity, by noticing that the model is successfully parsimonious and achieves a satisfactory clustering performance even with only a smaller subset of the informative variables.

%% \begin{figure}[!h]
%%    \centerline{\includegraphics[height=6cm, width=10cm, angle=0]{ROC.pdf}}
%%    \caption{Receiver Operating Characteristic, this plot illustrates the possible risk that by including too many variables we could also have many noise variables slipping through. In reality we observe that our model is fairly accurate as it can identify and include the greater majority of informative variables with a very small error rate.
%% \label{fig:roc}}
%%  \end{figure}

\section{Financial Markets Data}\label{sec:data}

\subsection{Applied Problem}

In the financial literature it is common practice to group markets into macro sectors
based on the type and nature of the good exchanged. Practitioners operating in
financial markets adhere to this convention and consider each sector as a separate
area of expertise. This approach is reasonable for fundamental investors who have
to be knowledgeable on the underlying factors driving demand/offer and have to
elaborate the relevant information as news become public.

A partition of the markets that mirrors the macro sectors is less 
suitable to systematic traders who take investment decisions based on algorithms
which depend only on the evolution of prices. Under these circumstances, developing
and optimizing a quantitative strategy on a sector by sector basis seems
rather arbitrary. This is because the only input considered when engineering the
strategy is the time series of prices whose behaviour is not necessarily a function
of the sector. A clustering method which is more consistent with a systematic and
objective approach, should identify homogeneous clusters of markets that share
similar price dynamics characteristics.

Our approach starts by selecting, across all sectors, those major financial markets
for which we have records spanning up to twenty years of trading (see the following Section). Under the
assumption that all the relevant information about a market can be extracted from
the historical prices, we then compute for each market the summary statistics that
measure the critical features of the distribution and the temporal dependence of
time series of returns.

In a supervised learning framework, the statistics of the price dynamics
can be seen as explanatory variables that can help us understand why the trading
performance is different across markets. When we apply a trading
algorithm to every market, one can observe that the risk-adjusted profit we obtain is
not consistent across markets. The supervised model we propose should be able to
regress the profitability of the trading algorithm on some of the features we record
for each market. An unsupervised approach is considered in Cozzini (2012), where
the relative benefits of supervised versus unsupervised learning are investigated.

Assuming we achieve a more accurate partition of the markets, we are in an
ideal position to develop a systematic trading strategy that better suits the markets
within each group. A strategy that has been optimized on a market by market
basis would likely be overfitted and would not have enough back testing (i.e.~for testing the algorithm) data. If,
instead, we devise a trading algorithm that consistently performs on a group of
markets, we are bound to obtain a more robust and convincing result. At the same
time, the significant features that are responsible for driving the clustering process
give us an insight on the critical aspects of price dynamics that should be exploited
by the trading strategy.

In order to reach credible conclusions about how to partition markets and what
are the informative features of the price dynamics, we need a clustering method
which is able to address the following issues characteristic of financial data:
\begin{itemize}
\item{Outliers and potential skewness}
\item{Non Informative Variables}
\item{Fewer Observations than Explanatory Variables.}
\end{itemize}
If we succeed in proposing a model whose performance is not hindered by these
issues, we will have increased confidence in the trading strategy that we develop
based on the outcome of clustering and variable selection process. By being able to
implement a more targeted strategy on each group of markets, we should achieve
better investment returns.

\subsection{The Specific Data Analyzed}\label{sec:design}

The dataset analysed has been kindly provided by AHL Research, a quantitative
investment manager, and integrates external sources with proprietary records of
live prices sampled during actual trading activity. The selection of markets considered
covers several sectors, assets classes and regions. The details of each market
considered are listed in Table \ref{tab:modelse}. The frequency of the samples is daily, typically
the end of day official settlement price, whenever the exchange provides one.

\begin{table}[!h] 
{\scriptsize
\begin{center}
\begin{tabular}{|c|c|c|c|c|c|c|}
\hline 
MARKET & SECTOR & DESCRIPTION & EXCHANGE & TYPE & START & CCY \\ 
\hline 
ADL    &  METAL     & Aluminium & LME & C                        & 19900101 & US \\ 
CPN    &  METAL     & Copper.NY & COMEX & F                      & 19900101 & US \\
GLN    &  METAL     & Gold & COMEX & F                           & 19900101 & US \\ 
SLN    &  METAL     & Silver & COMEX & F                         & 19900101 & US \\ 

WHC    &  AG        & Wheat & CBOT & F                           & 19900101 & US \\ 
SBC    &  AG        & Soyabeans & CBOT & F                       & 19900101 & US \\ 
SGN    &  AG        & Sugar & CSCE & F                           & 19900101 & US \\ 
CFN    &  AG        & Coffee.NY & CSCE & F                       & 19900101 & US \\
CCN    &  AG        & Cocoa.NY & CSCE & F                        & 19900101 & US \\

ADUS   &  CURRENCY   & Australian D Vs USD  & IB & X              & 19900101 & US \\
SFUS   &  CURRENCY   & Swiss Franc Vs USD & IB & X                & 19900101 & US \\
UKUS   &  CURRENCY   & British Pound Vs USD & IB & X              & 19900101 & US \\
EUYN   &  CURRENCY   & Japanese Yen Vs Euro & IB & X              & 19900101 & YN \\
CDUS   &  CURRENCY   & Canadian Dollar Vs USD & IB & X            & 19900101 & US \\

ESPC   &  STOCK     & E.mini.SP500.Future & CME & F              & 19900101 & US \\
FTL    &  STOCK     & FTSE & LIFFE & F                           & 19900101 & UK \\
DXF    &  STOCK     & Dax.Index & DTB & F                        & 19901123 & EU \\
NKS    &  STOCK     & Nikkei.225 & SIMEX & F                     & 19900101 & YN \\
ESTF   &  STOCK     & Euro.STOXX & DTB & F                       & 20000609 & EU \\
HSH    &  STOCK     & Hang.Seng & HKFE & F                       & 19900101 & HK \\ 
KIS    &  STOCK     & Korean.KOSPI200.Ind & KSE & F              & 20000920 & KW \\ 
TWS    &  STOCK     & Taiwan.MSCI.Ind & SIMEX & F                & 19970109 & US \\

TNC    &  BOND      & 10yr.T.Notes & CBOT & F                    & 19900101 & US \\
GTL    &  BOND      & Gilts & LIFFE & F                          & 19900101 & UK \\
DBF    &  BOND      & Euro.BUND & EUREX & F                      & 19900101 & EU \\
JBT    &  BOND      & Japanese.Bond & TSE & F                    & 19900101 & YN \\
ABS    &  BOND      & Ausi.10yr.Bond & SFE & F                   & 19900101 & AD \\

EDC    &  IRATE     & Eurodollar & CME & F                       & 19900101 & US \\ 
SSL    &  IRATE     & Short.Sterling & LIFFE & F                 & 19900101 & UK \\
EUL    &  IRATE     & Euribor & LIFFE & F                        & 19900101 & EU \\
EYT    &  IRATE     & Euroyen & TIFFE & F                        & 19900101 & YN \\
ARS    &  IRATE     & Ausi.T.Bills & SFE & F                     & 19900101 & AD \\

CLN    &  ENERGY     & Crude.Oil.NY & NYMEX & F                   & 19900101 & US \\
HON    &  ENERGY     & Heating.Oil & NYMEX & F                    & 19900101 & US \\
RBN    &  ENERGY     & RBOB.Gasoline & NYMEX & F                  & 19900101 & US \\
PTL    &  ENERGY     & Gas.Oil & IPE & F                          & 19900101 & US \\
\hline
\end{tabular}
\end{center}
\caption{Data. {\bf Market:} Code identifies market. {\bf Type:} Contract used for transaction. Cash, {\tt X}, exchange traded futures, {\tt F} or forwards {\tt C}. {\bf Start:} First date of daily records. {\bf CCY:} Currency. \label{tab:modelse}}
}
\end{table}

For each of the $n = 36$ financial markets we have data for, we compute $p = 13$ statistics which we arrange in a $36 \times 13$ data matrix. The complete list of 13 variables analyzed are reported in Table \ref{tab:expvars}; they comprise a variety of statistics, some descriptive of the distribution of returns such as kurtosis, some constructed to maintain and characterize the time-series structure of the data, such as autoregression order or more advanced as the Rescaled Range statistic (e.g.~Hurst (1951)). It is not known which, if any, statistics can explain the response detailed below.

%%%%%%%%%%%%%%%%%%%%%%%%%%%%%
 \begin{table}[h]
\begin{center}
\begin{tabular}{llll}
  \hline
Rank &   Variable &  Selection Freq. & Description\\
\hline
1 & stdev        &  Standard Deviation of daily returns \\
2 & skew         &  Skewness of daily returns \\
3 & kurtosis     &  Kurtosis of Daily Returns distribution \\ 
4 & tdof         &  Estimated degrees of freedom for a fitted t distribution \\
5 & xi           &  Estimated tail shape parameter for a fitted Generalised Pareto distribution\\
6 & arord        &  Estimated autoregression order \\
7 & autoq        &  Box-Pierce Q-statistic from autocorrelation coefficients\\
8 & box2         &  Ljung-Box test statistic at lag 2 \\
9 & vrt          &  Variance Ratio Test statistic \\
10 & whid        &  Fractal dimension estimate using Whittle's method \\
11 & rs          &  Rescaled Range Statistic \\
12 & wavH        &  Estimated Hurst exponent using waveletFit from fArma R package \\
13 & ghe         &  Generalised Hurst Exponent \\
\hline
\end{tabular}
\caption{ Explanatory Variables. \label{tab:expvars} }
\end{center}
\end{table}

%As a preliminary overview, in Figure \ref{fig:cormat}, we plot the correlation matrix of the different statistics computed over all markets. We can see that some variables are, as expected, highly correlated, for example kurtosis and robust kurtosis, or variance and interquantile range. 

%  \begin{figure}[!h]
%    \centerline{\includegraphics[height=12cm, width=14cm, angle=0]{PlotCorrMat.pdf}}
%    \caption{Correlation between different statistics computed on all markets. \label{fig:cormat}}
%  \end{figure}

The ultimate goal of the study is to find a more appropriate systematic trading strategy whose parameters can be robustly calibrated on clusters of similar markets. To verify that the markets' return features we have described are related to the trading strategies we are interested in, we compute a risk adjusted measure of investment performance as response variable. We adopt a simple moving average crossover to generate buy or sell signals and target a constant risk profile by scaling positions according to a rolling volatility measure. Given a time series of prices $\{p_{t}\}_{t=1} ^T$, and ${\tt EMA}_1 =p_1$, the exponential moving average at time $t>1$ is
\begin{equation*}
{\tt EMA}_t =  \alpha\,p_t + (1-\alpha) {\tt EMA}_{t-1}
\end{equation*}
where $\alpha$ represents the degree of exponential decay of the weights associated to older prices. The value of $\alpha$ determines the speed at which the exponential moving average reacts to a new recorded price and ultimately how close it tracks the price process. To generate our position signal we compute a fast ${\tt EMA^{(F)}}$ and a slow ${\tt EMA^{(S)}}$ by fixing $\alpha^{(F)}=\{0.03\}$ and $\alpha^{(S)}=\{0.01\}$ respectively. A buy signal is then generated every time the fast moving average crosses from below the slow moving average; conversely, a sell signal is given when it crosses from above. The position, ${\tt pos}$, is then held proportional to the difference between the two moving averages,
\begin{equation*}
{\tt pos}_t = ({\tt EMA_t^{(F)}} - {\tt EMA_t^{(S)}})/ {\tt vol_t} 
\end{equation*}
where ${\tt vol}$ is a measure of the rolling volatility (see Cozzini (2012) for the details on how it is computed). The point of scaling the position proportionally to the volatility of the price process, is to automatically adjust the risk of our exposure to the perceived uncertainty of the market. In practice, when the volatility increases we would scale down our positions.  The annualised Sharpe Ratio (Sharpe, 1966) measures the average return per unit of risk and it is computed from the sequence of the daily profits and losses $r_t = {\tt pos}_{1-t} \times (p_t - p_{t-1})$ 
\begin{equation*}
\frac{250/T \sum_{t=1}^{T} r_t} {\sqrt{250\,\mathbb{V}\textrm{ar}(r)}}
\end{equation*}
where $250$ is the number of working days in a year. This is used as our response variable.

\subsection{Data Analysis}

We now apply the penalised mixture of $t$ distributions to our financial markets data. We fit the Bayesian mixture of Lasso regressions, propose a new clustering of the financial markets and compare the results to the original macro sector partition. In the present implementation of the mixture of regressions we will assume the number of components is $K=2$. 
We remark that we have considered the analysis with $K>2$, but did not find any qualitative improvements in the data analysis. 
%Recall, the observed dependent variable $\bm{y}$ is the Sharpe ratio of the simple moving average crossover strategy described in Section \ref{sec:design}.

To fit the model we follow the sampling procedure described in Section \ref{sec:sampproc} and execute 10000 iterations of the PMCMC algorithm with adaptive resampling. The acceptance rate of the updates for $\tau$, $s$ were $0.23$, $0.24$ respectively, which are sensible values. We should remark that to deal with the label switching problem, common to many Bayesian mixtures, at each iteration we order and relabel the clusters according to the average Sharpe Ratio of each cluster.  
 
\begin{figure}[!h]
   \centerline{\includegraphics[height=10cm, width=12cm, angle=0]{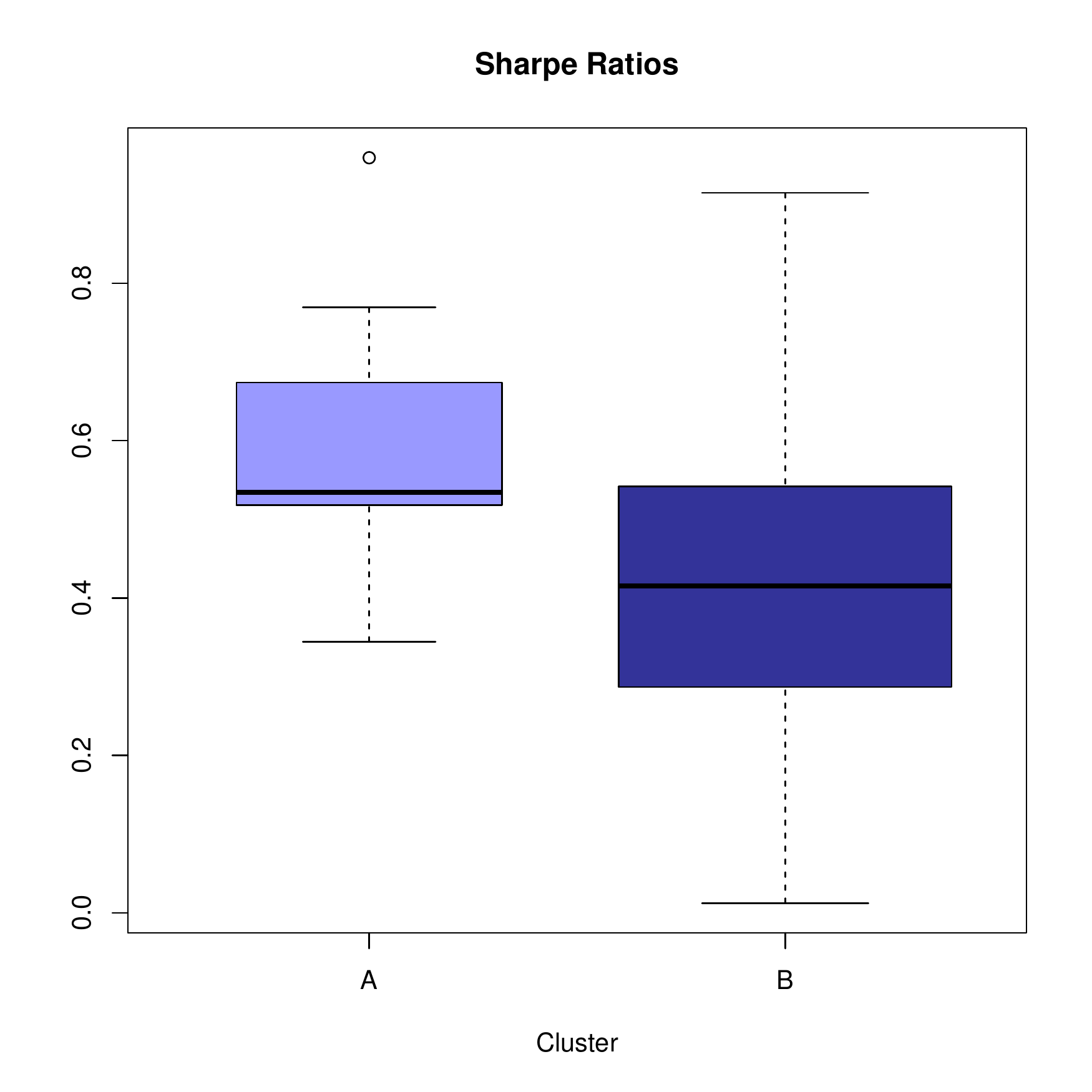}}
   \caption{Boxplots of Sharpe Ratios by cluster.\label{fig:boxplotsr}}
\end{figure}

In Table \ref{tab:freqvar} we quote the relative selection frequency of each variable across all PMCMC iterations. We compute the average of the $\bm{\gamma}$ indicator vector for each cluster over the 10000 iterations we run. According to the evidence, it appears that markets in cluster A are characterized by some specific features that are well represented by the Generalised Hurst Exponent (Di Matteo et al.~2004), the simple Hurst exponent computed using wavelet method and the Kurtosis. Since the markets in cluster A on average show a better Sharpe Ratio, we can conjecture an immediate link between the performance of the trend following strategies we tested and the persistence of markets as captured by these non-linear statistics as we can see in Figure \ref{fig:boxplotsr}. On the other hand, the boxplots on the right column of Figure \ref{fig:boxplotsr} seems to suggest that the markets in cluster B are characterized by a lower kurtosis and lower autocorrelation as measured by the Box-Pierce Q-statistic. 
%A more insightful explanation these results would require to investigate the exact value of the coefficients of the two regression curves. 

%%%%%%%%%%%%%%%%%%%%%%%%%%%%%
 \begin{table}[h]
\begin{center}
\begin{tabular}{ccc|ccc}
  \hline
\multicolumn{3}{c}{{Cluster A}} & \multicolumn{3}{c}{{Cluster B}} \\
Rank &   Variable &  Selection Freq. & Rank &   Variable &  Selection Freq.\\
\hline
1   & ghe       &  0.97    &  1   & kurtosis  &  0.66  \\
2   & wavH      &  0.72    &  2   & box2      &  0.57  \\
3   & kurtosis  &  0.7     &  3   & stdev     &  0.55  \\
4   & vrt       &  0.57    &  4   & tdof      &  0.53  \\
5   & tdof      &  0.57    &  5   & skew      &  0.52  \\
6   & xi        &  0.55    &  6   & autoq     &  0.49  \\
7   & stdev     &  0.53    &  7   & ghe       &  0.44  \\
8   & rs        &  0.49    &  8   & arord     &  0.44  \\
9   & whid      &  0.48    &  9   & xi        &  0.39  \\
10  & autoq     &  0.46    &  10  & vrt       &  0.37  \\
11  & box2      &  0.43    &  11  & whid      &  0.34  \\
12  & arord     &  0.41    &  12  & rs        &  0.33  \\
13  & skew      &  0.41    &  13  & wavH      &  0.31  \\
\hline
\end{tabular}
\caption{Ranking of the variables according to the frequency they have been selected to model Cluster A and Cluster B. \label{tab:freqvar} }
\end{center}
\end{table}

  \begin{figure}[!h]
   \centering
   \begin{minipage}[b]{7cm}
   %\centering
     \includegraphics[angle=0,height=6cm,width=6.5cm]  
                     {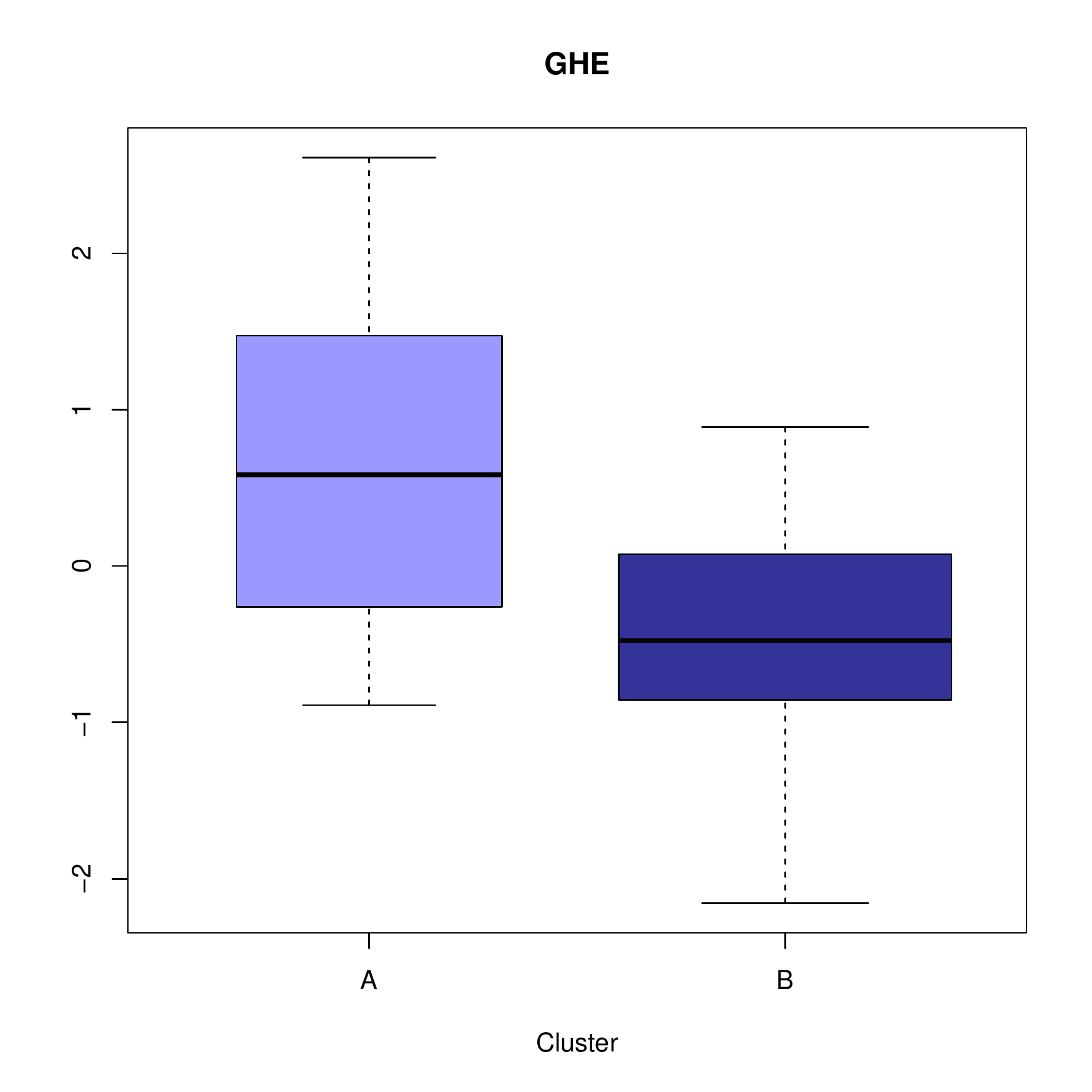}
    \end{minipage}
   \begin{minipage}[b]{7cm}
     \includegraphics[angle=0,height=6cm,width=6.5cm]  
                     {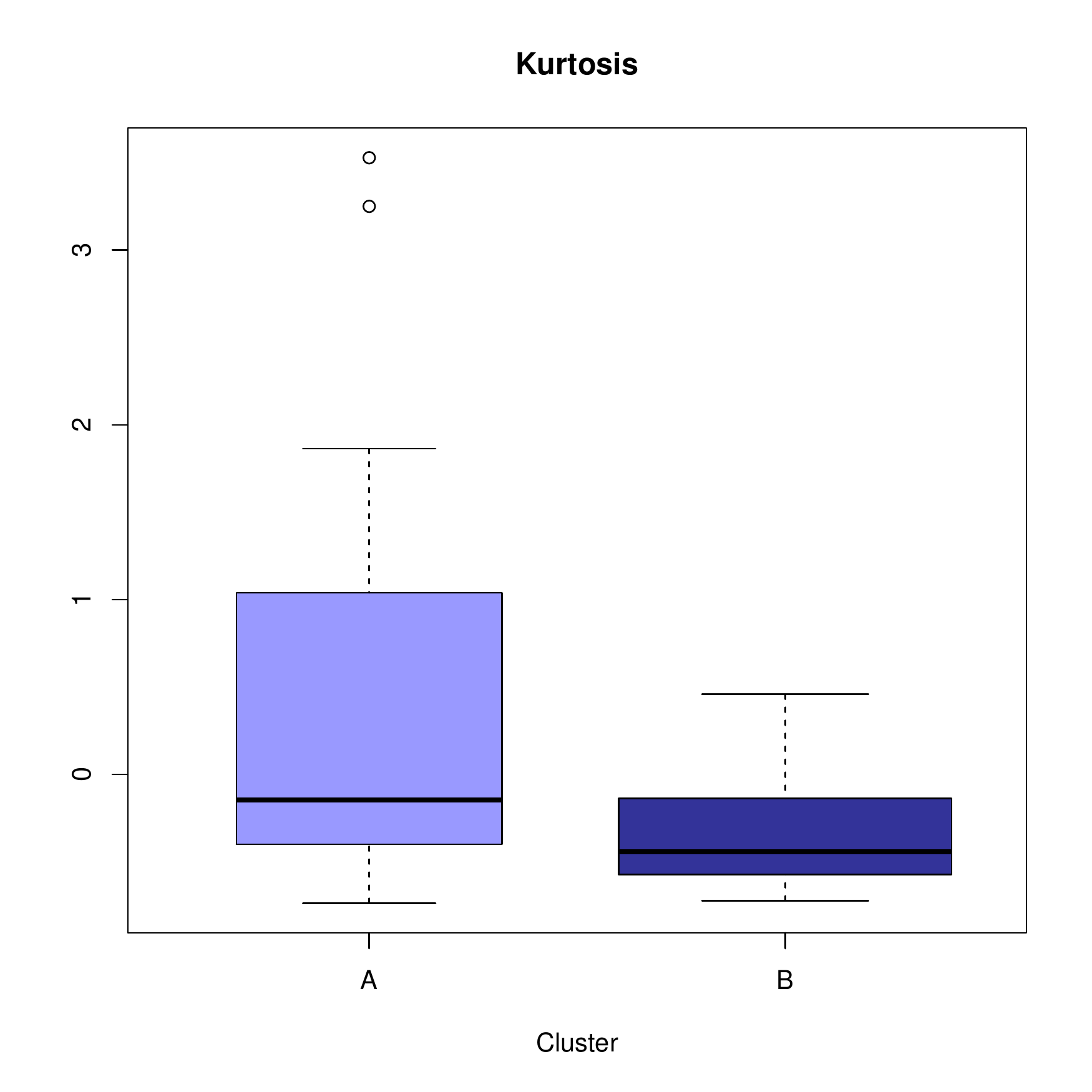}
   \end{minipage}
   \\[0mm]
  \begin{minipage}[b]{7cm}
     \includegraphics[angle=0,height=6cm,width=6.5cm]  
                     {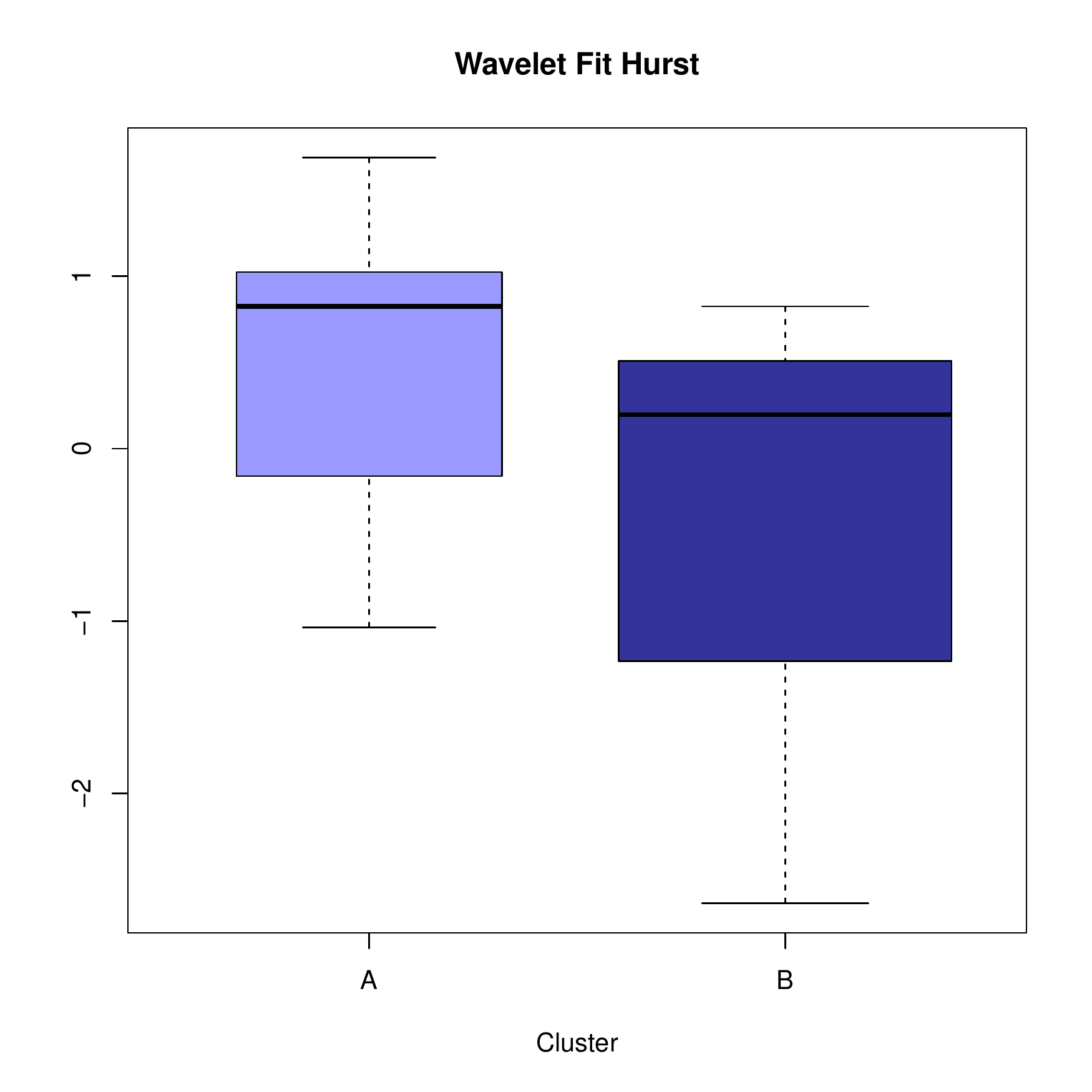}
    \end{minipage}
   \begin{minipage}[b]{7cm}
     \includegraphics[angle=0,height=6cm,width=6.5cm]  
                     {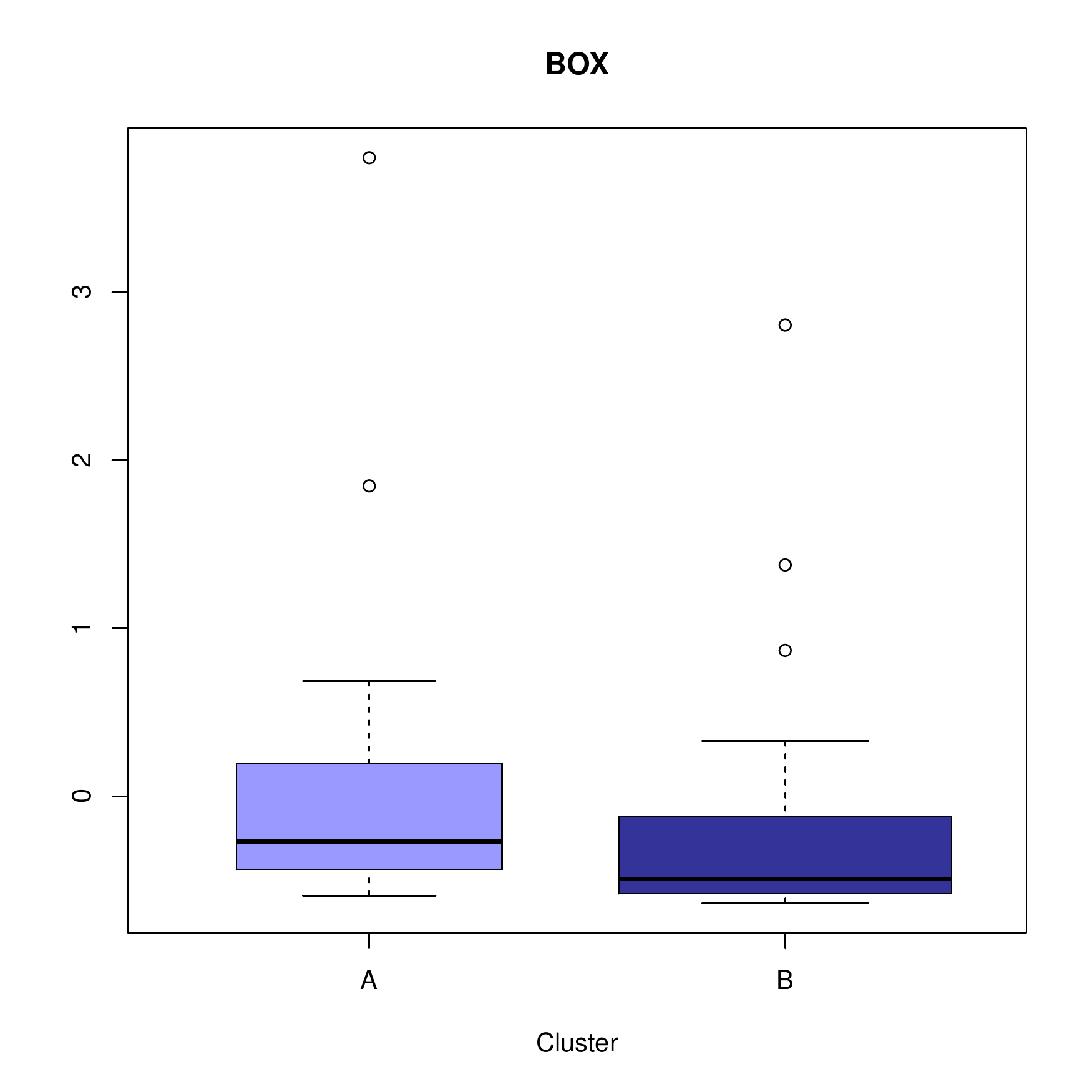}
    \end{minipage}
  \\[0mm]
     \caption{Boxplots of top ranking selected variables by cluster. \label{fig:boxplotvars}}
  \end{figure}

\begin{figure}[!h]
   \centerline{\includegraphics[height=14cm, width=16cm, angle=0]{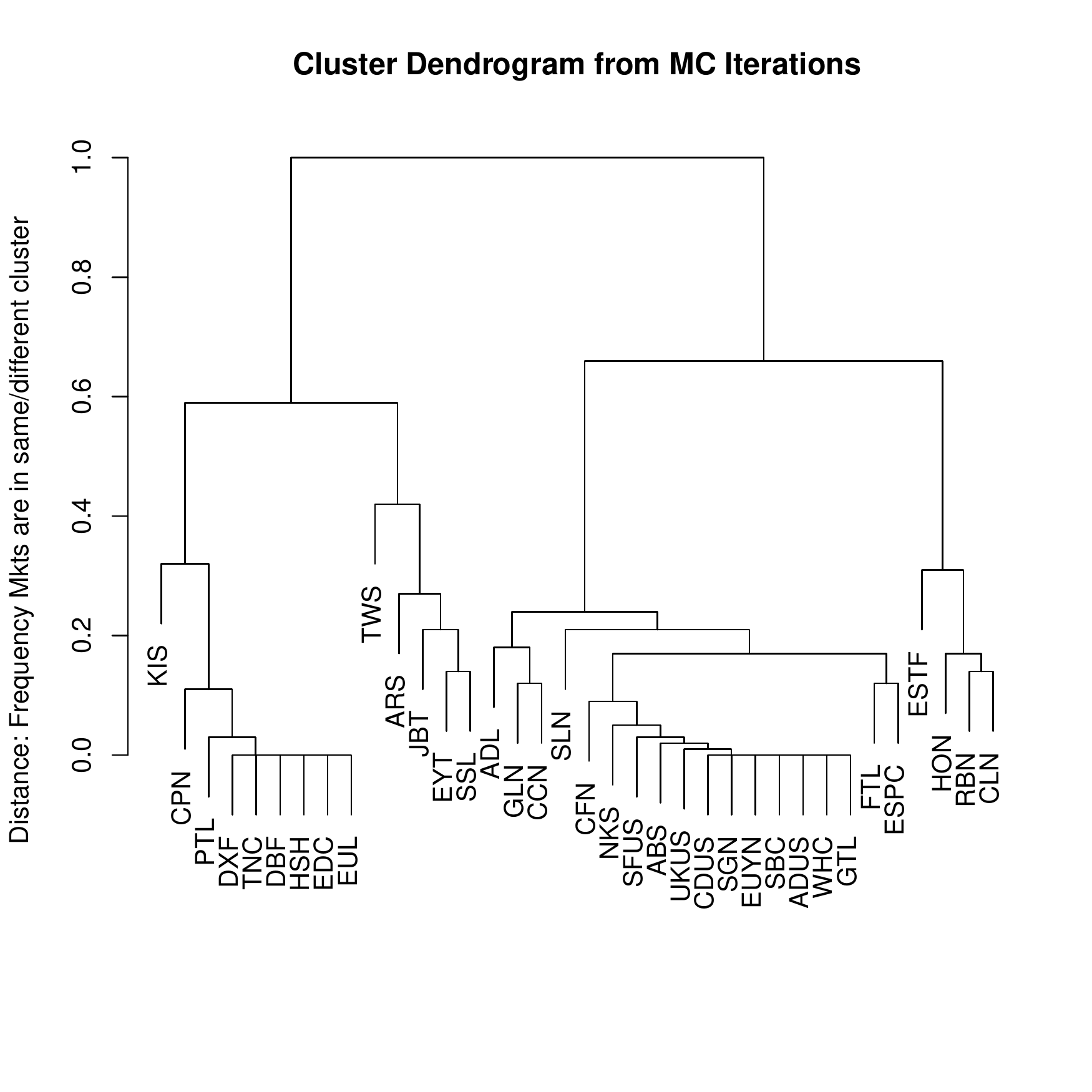}}
   \caption{Hierarchical clustering based on the relative frequency, over all PMCMC iterations, markets are assigned to the same clusters.\label{fig:dendro}}
\end{figure}

From the sampled posterior distribution of the label indicator we can infer how likely two markets belong to the same cluster. The relative frequency, over all PMCMC iterations, of the event that the two markets were assigned to the same cluster was used as a (posterior) measure that the two markets belong to the same cluster. 
Based on this, we can compute a distance that ranges between zero, if the markets are always in the same cluster, or one, if the markets are always assigned to distinct clusters. This approach allows us to compute a dissimilarity matrix between all markets and use this information to propose a hierarchical clusters as shown in Figure \ref{fig:dendro}.

%In particular we obtain that the measure will be one, if the markets are always in the same cluster,and zero if the markets are always assigned to distinct clusters. Looking at the relative frequency, over all PMCMC iterations, of the event where the two markets were assigned to the same cluster we can compute a probability measure. If the markets are always in the same cluster, the probability measure will be one, if the markets are always assigned to distinct clusters, the probability measure will be 0.

From the dendrogram in Figure \ref{fig:dendro} we can obtain a hard cluster assignment under the assumption $K=2$. The clustering results are reported in Table \ref{tab:clusmkts}. We can see how some markets that belong to the same macro sectors are kept together. For example, all interest rates markets are assigned to cluster A, while cluster B comprise all agriculturals and currency markets. Some other sectors are fairly evenly split between the two mixture components, this is the case for bond and stock markets. Still, in some cases, we can see a pattern where for example the more liquid stock markets are assigned to cluster B and the emerging markets, with the exception of German Dax, belong to cluster A.

 \begin{table}[h]
{\scriptsize
\begin{center}
\begin{tabular}{c|ccc}
  \hline
CLUSTER & MARKET &   SECTOR   &  SHARPE RATIO \\
\hline                        
A    &    JBT     &  BOND     &        0.52   \\
A    &    DBF     &  BOND     &        0.53   \\
A    &    TNC     &  BOND     &        0.53   \\
A    &    PTL     &  ENERGY   &        0.52   \\
A    &    EDC     &  IRATE    &        0.77   \\
A    &    ARS     &  IRATE    &        0.34   \\
A    &    EUL     &  IRATE    &        0.59   \\
A    &    EYT     &  IRATE    &        0.67   \\
A    &    SSL     &  IRATE    &        0.96   \\
A    &    CPN     &  METAL    &        0.59   \\
A    &    HSH     &  STOCK    &        0.53   \\
A    &    TWS     &  STOCK    &        0.73   \\
A    &    KIS     &  STOCK    &        0.35   \\
A    &    DXF     &  STOCK    &        0.42   \\
 \hline                    
B    &    WHC     &   AGS     &        0.56   \\
B    &    SBC     &   AGS     &        0.36   \\
B    &    SGN     &   AGS     &        0.36   \\
B    &    CCN     &   AGS     &        0.35   \\
B    &    CFN     &   AGS     &        0.45   \\
B    &    GTL     &  BOND     &        0.51   \\
B    &    ABS     &  BOND     &        0.7    \\
B    &    RBN     &  ENERGY   &        0.48   \\
B    &    CLN     &  ENERGY   &        0.9    \\
B    &    HON     &  ENERGY   &        0.91   \\
B    &    UKUS    &  CURRENCY &        0.54   \\
B    &    ADUS    &  CURRENCY &        0.22   \\
B    &    EUYN    &  CURRENCY &        0.37   \\
B    &    SFUS    &  CURRENCY &        0.57   \\
B    &    CDUS    &  CURRENCY &        0.26   \\
B    &    GLN     &  METAL    &        0.22   \\
B    &    ADL     &  METAL    &        0.5    \\
B    &    SLN     &  METAL    &        0.01   \\
B    &    NKS     &  STOCK    &        0.38   \\
B    &    ESTF    &  STOCK    &        0.29   \\
B    &    FTL     &  STOCK    &        0.52   \\
B    &    ESPC    &  STOCK    &        0.29   \\
\hline
\end{tabular}
\caption{Cluster assignment assuming $K=2$. \label{tab:clusmkts} }
\end{center}
}
\end{table}

As a further check of the relevance of the clustering proposed in Table \ref{tab:clusmkts} we can observe the value of the response variable $y$ for each cluster. In Figure \ref{fig:boxplotsr} we note a certain differentiation of the Sharpe Ratio between the two clusters, with cluster A markets generally showing a better performance of the trading strategy.

\section{Summary}\label{sec:summary}

We have considered a Bayesian mixture of lasso regressions with $t-$errors, designed for a financial data analysis problem. We applied a PMCMC algorithm to sample from a marginalized posterior and investigated the model and algorithm on simulated and real data.
In our real data example, we saw that the clusters returned by our model could perform better in terms of profitability on an in-sample basis, than the physical clustering of the market.
There are many issues that can be investigated in future work.

Firstly, with regards to the theoretical properties of the model. We did not investigate, for example, the issue of Lindley's paradox, which can manifest itself in mixtures (e.g.~Jennison (1997)). That is, we would like to know if there are some combinations of prior parameters, which would lead one to favouring statistical models with a single component. In connection to this, whether the complex posterior also satisfies a collection of inequalities for model probabilities as is the case for some standard Bayesian mixtures; see Nobile (2005). 

Secondly, the computational procedure of selecting the number of components. There are at least two options which we intend to consider in future work. The first is simply to use our PMCMC algorithm in each model. Then, as one can easily obtain a marginal likelihood estimate (indeed using the proposed particles -`all the samples' - see Andrieu et al.~(2010)) and compute Bayes factors - see e.g. Nobile (1994). The second idea is to build a trans-dimensional sampler based upon PMCMC and SMC samplers (Del Moral et al.~2006). Here, one uses a trans-dimensional version of the PMMH sampler. Suppose one has  a target density $\pi_k(x)$ in dimension $k$ and our overall target density is:
$
\pi(k,x) \propto \pi_k(x) p(k)\quad x\in\mathcal{X}^k\quad k\in\{1,\dots,k_{\textrm{max}}\}=\mathcal{K}
$
where $p(k)$ is a prior on the dimension (here the number of components in the mixture). Thus we have defined a target density on
$
\bigcup_{k\in\mathcal{K}} \{ k \} \times \mathcal{X}^k.
$
Now introduce a sequence of targets of dimension $k$:
$
\pi_{k,n}(x) \propto \pi_k(x)^{\gamma_n}
$
where $0<\gamma_1<\cdots<\gamma_p$ for some $p\geq 1$ given. Our trans-dimensional proposal is as follows: given a model order $k$, propose a model order $k'$ and use an SMC sampler to simulate the sequence $\pi_{k',n}$. The acceptance probability, when resampling at each time-point of the SMC algorithm, of such a move is:
$$
1\wedge \frac{\prod_{n=1}^p \frac{1}{N}\sum_{i=1}^N w_{n,k'}^i}{\prod_{n=1}^p \frac{1}{N}\sum_{i=1}^N w_{n,k}^i } \frac{p(k')q(k|k')}{p(k)q(k'|k)}
$$
where $ q(k'|k)$ is the proposal density of moving from $k$ to $k'$ and
$
\prod_{n=1}^p \frac{1}{N}\sum_{i=1}^N w_{n,k'}^i
$
is the marginal likelihood estimate from the SMC sampler in dimension $k'$. This allows one a possibility of producing very competitive trans-dimensional proposals.

\subsubsection*{Acknowledgements}
We thank AHL for providing the data. 
Work completed whilst the first author was PhD student at Imperial College London.
The second author was supported by an MOE grant.
We thank Nicolas Chopin and Dave Stephens for discussions on this work. We also
thank Elena Erlich and James Martin for their comments on the manuscript.

\vspace{0.025 in}

{\ \nocite{*} \centerline{ REFERENCES}
\begin{list}{}{\setlength{\itemindent}{-0.3in}}

\item
 \textsc{Andrieu}, C., \textsc{Doucet}, A. \& \textsc{Holenstein},
R.~(2010). Particle Markov chain Monte Carlo methods (with discussion).
\textit{J. R. Statist. Soc. Ser. B}, \textbf{72}, 269--342.

\item
{\sc Arnaud},  E. \& {\sc Le Gland}, F.~(2009).
SMC with adaptive resampling: large sample asymptotics, \emph{Proc. IEEE Workshop Statist. Sig. Proc.}.

\item
{\sc Cozzini}, A. M.~(2012).
\emph{Supervised and Unsupervised
Model-Based Clustering with Variable
Selection}. PhD.~Thesis, Imperial College London.

\item
{\sc Cozzini}, A. M., {\sc Jasra}, A. \& {\sc Montana} G. (2011). Robust model-based clustering with gene ranking,
Technical Report, Imperial College London.

\item
{\sc Del Moral}, P., {\sc Doucet}, A., \& {\sc Jasra}, A. (2006). Sequential Monte Carlo samplers.
\textit{J. R. Statist. Soc. Ser. B}
{\bf 68}, 411--436.

\item
{\sc Del Moral,} P., {\sc Doucet}, A. \& {\sc Jasra}, A.~(2012).
On adaptive resampling procedures for sequential Monte Carlo methods, {\it Bernoulli}, {\bf 18}, 252--278.

\item
{\sc Di Matteo}, T., { \sc Aste}, T., \& {\sc Dacorogna}, M. ~(2004).
Long term memories of developed and emerging markets: using the
  scaling analysis to characterize their stage of development.
\textit{ Journal of Banking \& Finance}, {\bf 4} 29:46.

\item
{\sc Diebolt}, J. \& {\sc Robert}, C. P. (1994).
Estimation of finite mixture distributions through Bayesian sampling.
\emph{J. R. Statist. Soc. Ser. B}, \textbf{56}, 363--375.

\item
{\sc Doucet}, A., {\sc Godsill}, S., \& {\sc Andrieu}, C. (2000). On sequential Monte Carlo sampling
methods for Bayesian filtering. \emph{Statist. Comp.} 197--208.

%\item
%{\sc Fahrmeir},  L., {\sc Kneib}, T., \& {\sc Konrath}, S. (2009). Bayesian regularisation in structured
%additive regression: a unifying perspective on shrinkage, smoothing and
%predictor selection. \emph{Statist. Comp.}, {\bf 20}, 203--219.

\item
{\sc Fearnhead}, P. \& {\sc Meligkotsidou}, L. (2007). Filtering Methods for Mixture Models.
\emph{J. Comp. Graph.l Statist.}, {\bf 16}, 586--607.

\item
{\sc George}, E. I. \& {\sc Mcculloch}, R. E. (1997). Approaches for Bayesian variable
selection. \emph{Statistica Sinica}, 7, 339--373.

\item
{\sc Goldfeld}, S. \& {\sc Quandt}, R. E. (1973). A markov model for switching regression.
\emph{J. Econom.}, {\bf 1},  3--15.

\item
{\sc Hurn}, M., {\sc Justel}, A., \& {\sc Robert}, C. P. (2003). Estimating mixtures of regressions.
\emph{J. Comp. Graph. Statist.}, {\bf 12}, 55--79.

\item
{\sc Hurst}, H.~E. (1951). Long Term Storage Capacity of Reservoirs.
\emph {Trans Am Soc Civil Eng.}, {\bf 116}, 770--808.

\item
{\sc Jasra}, A., {\sc Holmes}, C. C.~\& {\sc Stephens}, D. A.~(2005). Markov chain Monte Carlo
and the label switching problem in Bayesian mixture modelling. \textit{Statist. Sci.},
{\bf 20}, 50--67.

\item
{\sc Jennison}, C.~(1997). Discussion of On Bayesian analysis of mixtures with an unknown number
of components. \textit{J. R. Statist. Soc. Ser. B},  {\bf 59}, 731--792.

\item
{\sc Khalili}, A. \& {\sc Chen}, J. (2007). 
Variable selection in finite mixture of regression models. 
\emph{J. Amer. Statist. Assoc.}, {\bf 102}, 1025--1038. 

\item
{\sc Kim}, S., {\sc Tadesse}, M. G., \& {\sc Vannucci}, M. (2006). Variable selection in clustering
via Dirichlet process mixture models. \emph{Biometrika}, {\bf 93}, 877--893.

\item
{\sc Lo}, A. W. \& {\sc MacKinlay}, A. C. (1988). Stock market prices do not follow random
walks: evidence from a simple specification test. \emph{Rev. Finan. Stud.},
{\bf 1}, 41--66.

\item
{\sc Mukhopadhyay}, S. \& {\sc Bhattacharya}, S. (2011). Perfect Simulation for Mixtures
with Known and Unknown Number of components. arxiv preprint.

\item
{\sc Nobile}, A. (1994).\emph{ Bayesian Analysis of Finite Mixture Distributions.} PhD thesis,
Carnegie Mellon University.

\item
{\sc Nobile}, A. (2005). On the posterior distribution of the number of components in
a finite mixture. \emph{Ann. Statist.}, {\bf 32}, 2044--2073.

\item
{\sc Park}, T. \& {\sc Casella}, G. (2008). The Bayesian lasso.
\emph{J. Amer. Statist. Assoc.}, {\bf 103}, 681-686.

\item
{\sc Raftery}, A. \& {\sc Dean}, N. (2006). Variable selection for model-based clustering.\emph{J. Amer. Statist. Assoc.}, {\bf 101}, 168--178

\item
{\sc Robert}, C. P. \& {\sc Casella}, G. (2004). \emph{Monte Carlo Statistical Methods}.
Springer: New York.

\item
{\sc Sch\"afer}, C. \& {\sc Chopin}, N. (2012). 
Adaptive Monte Carlo on binary sampling spaces. \emph{Statist. Comp.} (to appear).

\item
{\sc Sharpe}, W. F. (1966). Mutual fund performance. \emph{J. Business},
{\bf 39}, 119--138.

\item
{\sc Tibshirani}, R.~(1996).  Regression selection and shrinkage via the lasso.
\emph{J. R. Statist. Soc. Ser. B}, {\bf 58}, 267--288.

\item
{\sc Yau}, C. \& {\sc Holmes}, C. C.~(2011). Hierarchical Bayesian nonparametric mixture
models for clustering with variable relevance determination. \emph{Bayes. Anal.}, {\bf 6}, 329--352.

\end{list}
}

\end{document}